\documentclass[useAMS,usenatbib]{mn2e}
\usepackage{graphicx}
\usepackage{psfig}
\usepackage{amssymb,amsmath}
\usepackage{mathrsfs}
\usepackage{times}
\usepackage{setspace}
\usepackage{color}

\def\aj{AJ}%
\def\araa{ARA\&A}%
\def\apj{ApJ}%
\def\apjl{ApJ}%
\def\apjs{ApJS}%
\def\aap{A\&A}%
\def\mnras{MNRAS}%
\def\nat{Nature}%
\newcommand{\begit}{\begin{itemize}}
\newcommand{\enit}{\end{itemize}}
\newcommand{\begen}{\begin{enumerate}}
\newcommand{\enen}{\end{enumerate}}
\newcommand{\p}{\partial}    
 
\newcommand{\beq}{\begin{equation}}
\newcommand{\eeq}{\end{equation}}
\newcommand{\beqa}{\begin{eqnarray}} 
\newcommand{\eeqa}{\end{eqnarray}} 
\def\lesssim{\mathrel{\hbox{\rlap{\hbox{\lower5pt\hbox{$\sim$}}}\hbox{$<$}}}}
\def\gtrsim{\mathrel{\hbox{\rlap{\hbox{\lower5pt\hbox{$\sim$}}}\hbox{$>$}}}}

\title[Multi-Phase gas in Galactic Winds and Halos]
{An Origin for Multi-Phase Gas in Galactic Winds and Halos}

\author[Thompson et al.]{
Todd A.~Thompson$^1$, Eliot Quataert$^2$, Dong Zhang$^1$, \& David H.~Weinberg$^1$\\
$^1$Department of Astronomy and Center for Cosmology \& Astro-Particle Physics, The Ohio State University, Columbus, Ohio 43210\\
$^2$Departments of Physics and Astronomy, Theoretical Astrophysics Center, UC Berkeley, Berkeley, CA 94720}

\begin{document}
\maketitle
\label{firstpage}
\begin{abstract}

The physical origin of high velocity cool gas seen in galactic winds remains unknown. Following \cite{wang95a}, we argue that radiative cooling in initially hot thermally-driven outflows can produce fast neutral atomic and photoionized cool gas.  The inevitability of adiabatic cooling from the flow's initial $10^7-10^8$\,K temperature and the shape of the cooling function for $T\lesssim10^{7}$\,K imply that outflows with hot gas mass-loss rate relative to star formation rate of  $\beta=\dot{M}_{\rm hot}/\dot{M}_\star\gtrsim0.5$ cool radiatively on scales ranging from the size of the energy injection region to tens of kpc.  We highlight the $\beta$ and star formation rate surface density dependence of the column density, emission measure, radiative efficiency, and velocity. At $r_{\rm cool}$, the gas produces X-ray and then UV/optical line emission with a total power bounded by $\sim10^{-2}\,L_\star$ if the flow is powered by steady-state star formation with luminosity $L_\star$. The wind is thermally unstable at $r_{\rm cool}$, potentially leading to a multi-phase medium. Cooled winds decelerate significantly in the extended gravitational potential of galaxies. The cool gas precipitated from hot outflows may explain its prevalence in galactic halos. We forward a picture of winds whereby cool clouds are initially accelerated by the ram pressure of the hot flow, but are rapidly shredded by hydrodynamical instabilities, thereby increasing $\beta$, seeding radiative and thermal instability, and cool gas rebirth. If the cooled wind shocks as it sweeps up the circumgalactic medium, its cooling time is short, thus depositing cool gas far out into the halo. Finally, conduction can dominate energy transport in low-$\beta$ hot winds, leading to flatter temperature profiles than otherwise expected, potentially consistent with X-ray observations of some starbursts.

\end{abstract}


\section{Introduction}
\label{section:introduction}

Galactic winds are crucial to the process of galaxy formation, ejecting gas from galaxies \citep{ham}, helping to regulate star formation, shaping the stellar mass function and the mass-metallicity relation \citep{dekel,finlator_dave,erb08,peeples_shankar}, and enriching  the intergalactic medium with metals \citep{aguirre01,oppenheimer_dave06,oppenheimer_dave08}.  Hot winds heated by supernovae and stellar winds or active galactic nuclei may be present in both star-forming and passive galaxies at all redshifts.  Quantifying their prevalence, dynamical importance, and observational signatures is a key area in observational and theoretical studies of galactic winds.  

An important outstanding question is the nature of the cool clouds of molecular, neutral atomic, and ionized gas seen in blue-shifted absorption and emission in rapidly star-forming galaxies and starbursts at all redshifts.  The prevailing picture is that these clouds are driven out of the host galaxy by ram pressure acceleration from a supernova-heated hot wind (e.g., \citealt{veilleux_review}), but it is unclear if clouds can be accelerated to the required velocities before being shocked and shredded by hydrodynamical instabilities \citep{scannapieco,zhang15}.  Additional mechanisms for cold cloud acceleration have also been suggested, including momentum deposition by supernovae, the radiation pressure of starlight on dust grains \citep{mqt,mqt10,mmt,hopkins12_wind,zhang_thompson,krumholz_thompson13,davis_kt,thompson_krumholz,thompson_2015}, and cosmic rays \citep{breitschwerdt94,breitschwerdt99,jubelgas,socrates_cr,hanasz}.

Here, we revisit the physics of the radiative cooling of hot, initially adiabatic flows. We argue for a picture where the cool gas in galactic winds and halos precipitates directly from the hot wind as a result of radiative cooling, based on earlier work by \cite{wang95a,wang95b}, which was subsequently developed by \cite{efstathiou2000} and \cite{silich2003,silich2004,tenorio,tenorio05,tenorio07,wunsch}, and suggested in ultra-luminous infrared galaxies by \cite{martin_lya}.

In Section \ref{section:analytic} we present an analytic discussion that shows that mass-loaded winds always cool on large scales. The key point is that adiabatic cooling lowers the gas temperature to a value where further decreasing the temperature leads to more rapid cooling (e.g., Fig.~\ref{figure:c}). In this way, the cooling timescale can become shorter than the dynamical time for expansion, the assumption of adiabaticity breaks down, and the flow rapidly cools, radiating its remaining thermal energy content. We provide simple scalings that give the cooling radius, radiative efficiency, emission measure, column density, and velocity at the cooling radius. We further identify critical values of the star formation rate surface density and mass-loading rate for the cooling instability.  Section \ref{section:conduction} shows when we expect conduction to dominate energy transport in hot winds, leading to much flatter temperature profiles close to the central galaxy. 

In Section \ref{section:numerical} we solve the general equations for a time-steady flow with an arbitrary cooling/heating function. The strongly cooled wind forms an extended region of cool photoionized outflowing gas. We show that the extended gravitational potential of galaxies decelerates the cooled wind on $\sim{\rm few}-100$\,kpc scales.  Section \ref{section:ti} discusses the linear instability of the cooling flow.  We show that it is both convectively and thermally unstable, and that the latter should enhance density fluctuations by of order $\sim100$.  In Section \ref{section:discussion} we discuss our results in terms of observations of systems including NGC 253 and M82, ultra-luminous infrared galaxies with high velocity line emission, high redshift rapidly star forming galaxies, and the absorption line systems both at $z\sim0$ observed by COS-Halos \citep{werk}, and at high-redshift \citep{steidel_cgm}. We note in Section \ref{section:rebirth} that the very strong dependence of the cooling radius and efficiency on the mass-loading rate $\beta$ motivates a picture of galactic winds where the cool clouds are initially launched by ram pressure acceleration and are rapidly destroyed by hydrodynamical instabilities. This process seeds the hot flow with both a higher mass loading rate, that rapidly causes cooling on larger scales, and density fluctuations that grow rapidly as the thermal instability develops. In Section \ref{section:halo} we discuss the origin of the cool gas in halos and we calculate the cooling time of the wind after it shocks on the surrounding circumgalactic medium while blowing a wind-driven bubble. Our conclusions are presented in Section \ref{section:conclusions}.

\section{Wind Cooling}
\label{section:winds}

\subsection{Analytic Considerations}
\label{section:analytic}

One might expect that a hot spherical wind never cools radiatively since it rapidly expands to low density ($n\propto r^{-2}$).  To see that it can, first note that adiabatic cooling of an ideal gas dictates that $P/\rho^\gamma={\rm constant}$, which implies that $T\propto r^{-4/3}$ ($\gamma=5/3$).  As the super-heated flow initially expands from $T\sim10^8$\,K, cooling is dominated by bremsstrahlung, the cooling time is $t_{\rm cool}\propto T^{1/2}/n\propto r^{4/3}$, the advection time is $t_{\rm adv}\sim r/v\propto r$, so that the ratio $t_{\rm cool}/t_{\rm adv}\propto r^{1/3}$ grows slowly with radius.  Thus, if the flow starts with $t_{\rm cool}/t_{\rm adv}>1$, it remains adiabatic as long as bremsstrahlung cooling dominates cooling.

\begin{figure}
\centerline{\includegraphics[width=8.5cm]{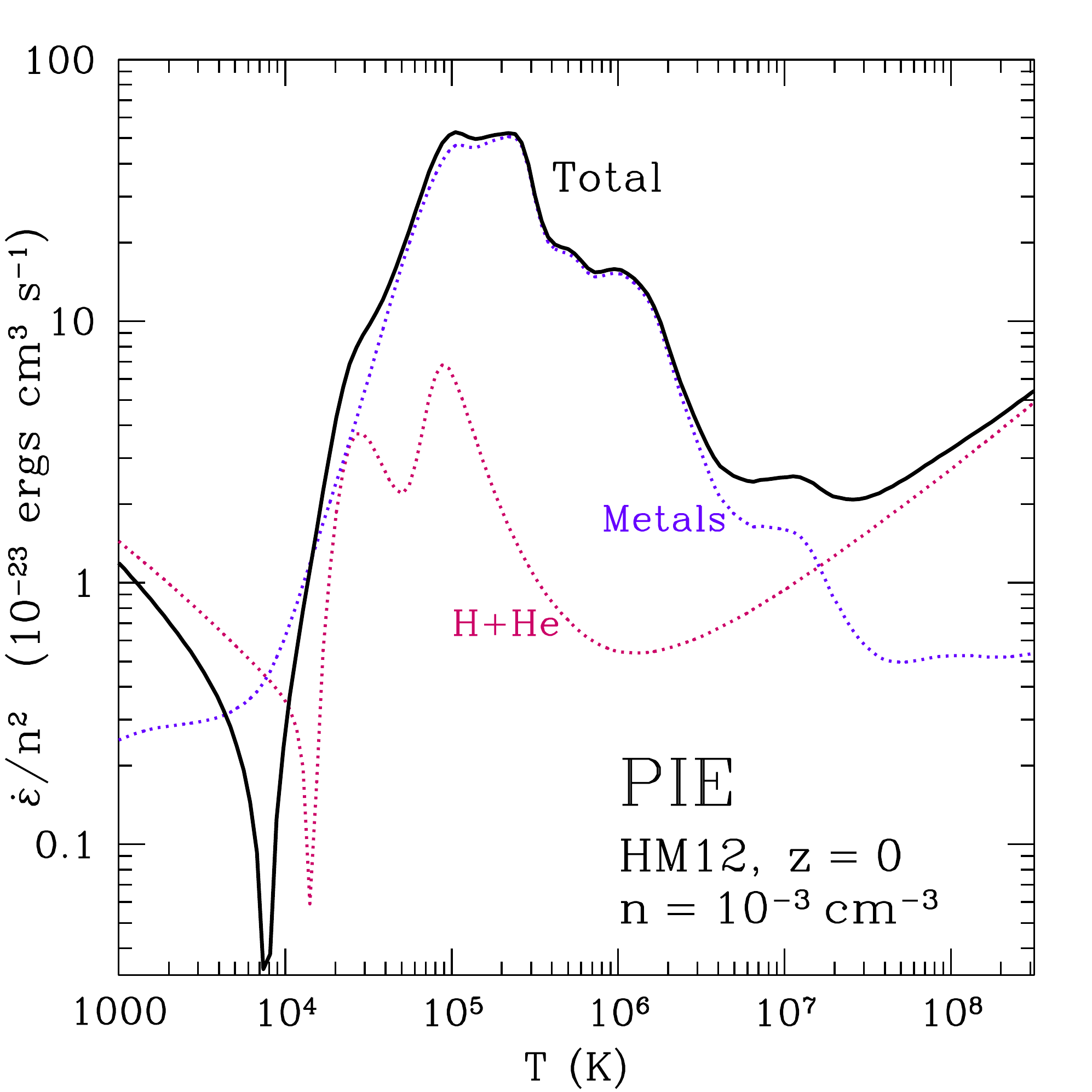}}
\caption{Illustrative example of the cooling/heating function in photoionization equilibrium (PIE) with a metagalactic UV background \citep{oppenheimer_schaye,wiersma}.  The red line shows the contribution from H and He, whereas the purple line shows the contribution from metals.  Bremsstrahlung dominates cooling at high temperatures. Heating dominates at low temperatures (the ``spike" at $\sim10^4$\,K shows where $|\dot{\varepsilon}|$ goes through zero). Equation (\ref{cool}) gives a simple approximation to the cooling function between $10^5\lesssim T\lesssim10^7$\,K for solar-metallicity gas in collisional ionization equilibrium (CIE) and over a range of gas densities in PIE.}
\label{figure:c}
\end{figure}

The situation changes when the flow cools enough that metal lines dominate the gas emissivity at $T\lesssim10^7$\,K.  At these lower temperatures, which the  flow inevitably reaches at sufficiently large distance as a result of adiabatic cooling, the shape of the cooling function changes, and can be approximated as  (e.g., \citealt{maclow,draine}) 
\beq
\Lambda(T)=\dot{\varepsilon}/n^2\simeq3\times10^{-23}\,T_7^{-0.7}\xi\,\,\,(10^5\lesssim T\lesssim10^7\,{\rm K}),
\label{cool}
\eeq
where $T_7=T/10^7$\,K and where $\xi$ is meant to capture the dependence of $\Lambda$ on the metallicity of the gas: $\xi=1$ for Solar composition and $\xi\sim0.3$ and $\sim3$ for composition of $0.1$ and $3$ times Solar, respectively. In this regime $t_{\rm cool}\propto r^{-4/15}$ and the ratio $t_{\rm cool}/t_{\rm adv}\propto r^{-19/15}$ decreases rapidly with radius.  Thus, even if the flow starts with $t_{\rm cool}/t_{\rm adv}>1$, on large physical scales it has a chance to become radiative.  Although the temperature scaling, normalization, and density dependence of the cooling function depends on the metallicity of the medium and whether or not it is collisionally ionized, photoionized, or in ionization equilibrium, the upturn in $\Lambda(T)$ at low temperatures is unavoidable because of line cooling, even in a gas of primordial composition.  An example of a representative cooling function for solar metallicity gas in photoionization equilibrium (PIE) with the meta-galactic UV background is shown in Figure \ref{figure:c}.  Note that the outflows powered by supernovae in rapidly star-forming galaxies may have super-solar metallicities, depending on the mass loading and metallicity of the surrounding ISM. More discussion of metallicity in the context of radiative cooling is provided in Section \ref{section:numerical}.

As in \cite{wang95a,wang95b} (see also \citealt{efstathiou2000,silich2003,silich2004}), we adopt the wind model of \cite{cc85} (hereafter CC85) to evaluate the importance of radiative cooling. We define dimensionless parameters $\beta$ and $\alpha$ such that  
\beq
\dot{M}_{\rm hot}=\beta\,{\dot{M}_\star}
\label{mdothot}
\eeq 
and  
\beq
\dot{E}_{\rm hot}=3\times10^{41}\,{\rm ergs\,\,s^{-1}}\,\alpha\,{\dot{M}_\star}, 
\label{edothot}
\eeq
where the latter assumes that $\alpha10^{51}$\,ergs per core-collapse supernova (SN) is thermalized in the hot plasma, that there are $10^{-2}$\,${\rm SNe/yr}$ per M$_\odot/{\rm yr}$ of star formation, and that the star formation rate $\dot{M}_\star$  is measured in M$_\odot/{\rm yr}$.  For a spherical galaxy of size $R$, the CC85 model assumes a constant mass loading rate and energy deposition rate per unit volume for $r<R$ and that the flow is purely adiabatic and mass-conserving for $r>R$.  Given the parameters $\alpha$ and $\beta$, the solution then provides the radial run of pressure, density, and velocity as a function of radius at all $r$.  The character of the solution is such that the Mach number ${\cal M}=1$ at $R$. Appendix \ref{appendix:cc85} provides a discussion of the CC85 model.

Using the CC85 wind model for a wind expanding into a fixed solid angle $\Omega$, employing the cooling function given in equation (\ref{cool}), and assuming that $r\gg R$ ($v\approx$\,constant), we find that 
\beq
t_{\rm cool}\simeq3\times10^{6}\,{\rm yr}\frac{\alpha^{2.20}}{\beta^{3.20}} \left(\frac{R_{0.3}}{r_{10}}\right)^{0.27}\frac{R_{0.3}^{2}}{{\dot{M}_{\star,10}}}\,\frac{\Omega_{4\pi}}{\xi},
\label{tcool}
\eeq
and
\beq
t_{\rm adv}\simeq1\times10^{7}\,{\rm yr}\,\left(\frac{\beta}{\alpha}\right)^{1/2}\,r_{10},
\label{tadv}
\eeq
where $R_{0.3}=R/0.3$\,kpc, $r_{10}=r/10$\,kpc, $\dot{M}_{\star,10}=\dot{M}_\star/10$\,M$_\odot$ yr$^{-1}$, $\Omega_{4\pi}=\Omega/4\pi$, and we have dropped the dependence on the mean molecular weight and Hydrogen fraction in $t_{\rm cool}$ (we assume $\mu=1$ throughout this paper). In equation (\ref{tcool}) for $t_{\rm cool}$, note the importance of the star formation rate surface density $\dot{\Sigma}_\star={\dot{M}_\star}/\pi R^2$, which varies among rapidly star-forming galaxies by more than four orders of magnitude \citep{kennicutt98}.  Systems with higher $\dot{\Sigma}_\star$ have shorter flow cooling timescale \citep{strickland_heckman}.

The choice of parameters for scaling the relations above (and below) is motivated in part by M82, which has $R\simeq250$\,pc and $\dot{M}_\star\simeq5-10$\,M$_\odot$ yr$^{-1}$ (e.g., \citealt{rieke,forster2001,forster2003}).  In addition, from X-ray observations M82 has been inferred to have $\alpha\simeq1$ and $\beta\simeq0.2-0.5$ \citep{strickland_heckman}. These values of $\dot{M}_\star$ and $R$ are also roughly appropriate for the individual massive star-forming clumps seen in high redshift galaxies \citep{genzel_clump}. We will see below that the propensity for radiative cooling and its radiative efficiency are strong functions of the host galaxy parameters. The applicability to individual systems and collections of systems is discussed in Section \ref{section:discussion}.

Setting $t_{\rm cool}=t_{\rm adv}$, we derive the critical radius beyond which the flow becomes radiative in the $r\gg R$ limit:
\beq
r_{\rm cool}\simeq4\,{\rm kpc}\,\frac{\alpha^{2.13}}{\beta^{2.92}}\,R_{0.3}^{1.79}\,\left(\frac{\Omega_{4\pi}}{\xi\dot{M}_{\star,10}}\right)^{0.789}.
\label{rcool}
\eeq
Note the very strong dependence on both $\alpha$ and $\beta$.  A small change in either can move $r_{\rm cool}$ significantly.  However, $\beta$ cannot be arbitrarily small and still have rapid cooling.  Because the cooling function has a peak and drops at temperatures below $\sim10^5$\,K (e.g., Fig.~\ref{figure:c}), and because  $t_{\rm cool}/t_{\rm adv}$ decreases as $r^{-19/15}$ while the cooling function of equation (\ref{cool}) is applicable, there is a minimum value of the mass loading parameter $\beta$ such that rapid radiative cooling occurs anywhere in the wind:
\beq
\beta_{\rm min}\simeq0.64\,\alpha^{0.636}\,\left(\frac{R_{0.3}}{\dot{M}_{\star,10}}\frac{\Omega_{4\pi}}{\xi}\right)^{0.364}.
\label{betamin}
\eeq
For $\beta\lesssim\beta_{\rm min}$, the flow remains adiabatic to arbitrarily large scales, and although  $t_{\rm cool}/t_{\rm adv}$ decreases in the temperature range where equation (\ref{cool}) is applicable, it never crosses unity before the cooling function turns over at temperatures below $\sim10^5$\,K (Fig.~\ref{figure:c}). For a ULIRG with $\dot{M}_\star\sim100$\,M$_\odot$ yr$^{-1}$, $\beta_{\rm min}\simeq0.3$. For an LBG with larger $R\sim\,{\rm kpc}$ and $\dot{M}_\star\sim30$\,M$_\odot$ yr$^{-1}$, $\beta_{\rm min}\simeq0.6$. These scalings imply that radiative cooling of hot galactic winds may be prevalent both in local starbursts and in galaxies on the star forming main sequence at high redshift \citep{wuyts11}.

Equation (\ref{rcool}) can be re-written as a critical condition on $\beta$, such that $r_{\rm cool}$ obtains at a radius $r\gg R$
\beq
\beta_{\rm crit}\simeq0.73\alpha^{0.730} \left(\frac{10\,{\rm kpc}}{r_{\rm cool}}\right)^{0.342}R_{\rm 0.3}^{0.613}
\left(\frac{\Omega_{4\pi}}{\xi\dot{M}_{\star,10}}\right)^{0.270}.
\label{betacrit1}
\eeq  
The fact that $\beta_{\rm crit}$ is similar to $\beta_{\rm min}$ implies that the maximum value of the cooling radius is $\sim10$\,kpc for our nominal scaling parameters. Substituting, we find that the maximum value of the cooling radius is 
\beq
r_{\rm cool}^{\rm max}(\beta=\beta_{\rm min})\simeq15\,{\rm kpc}\,\alpha^{0.274}R_{0.3}^{0.728}\left(\frac{\xi\dot{M}_{\star,10}}{\Omega_{4\pi}}\right)^{0.273}.
\eeq
If the wind is mass-loaded sufficiently to cool radiatively ($\beta>\beta_{\rm min}$), it does so on a scale smaller than $r_{\rm cool}^{\rm max}$.

By setting $r_{\rm cool}=R$ we derive the critical mass loading rate such that the cooling radius collapses to the scale of the launching radius $R$. For this calculation we use the general CC85 solution valid at $r=R$ (eq.~\ref{rcool} requires $r\gg R$).  We find that
\beq
\beta_{\rm crit}(r_{\rm cool}=R)\simeq1.95\alpha^{0.730}\left(\frac{\Omega_{4\pi}R_{\rm 0.3}}{{\xi \dot{M}_{\star,10}}}\right)^{0.270}.
\label{betacrit}
\eeq
The expressions above can be inverted to yield a critical star formation rate surface density ($\dot{\Sigma}_\star={\dot{M}_\star}/\pi R^2$) for the flow to cool at any given radius $r\gg R$
\beq
\dot{\Sigma}_{\star,\,\rm crit}\simeq 3.5\,\frac{{\rm M_\odot}}{\rm yr\,kpc^{2}}\,R^{0.267}_{0.3}\left(\frac{\rm 10\,kpc}{r_{\rm cool}}\right)^{1.27}\frac{\alpha^{2.70}}{\beta^{3.70}}\frac{\Omega_{4\pi}}{\xi},
\label{sdscrit}
\eeq
which is exceeded by starbursts in the local universe and by rapidly star-forming galaxies at high-redshift (e.g., \citealt{wuyts11}). 

We can also estimate the critical star formation surface density required for the flow to cool at $r_{\rm cool}=R$ using equation (\ref{betacrit}), which employs the full solution to the CC85 equations at $r=R$ and not the limit $r\gg R$ (see Appendix A; see also Appendix A of \citealt{strickland_heckman}). We find that 
\beq
\dot{\Sigma}_{\star,\,\rm crit}(r_{\rm cool}=R)\simeq420\,\frac{{\rm M_\odot}}{\rm yr\,kpc^{2}}\,\frac{\alpha^{2.70}}{\beta^{3.70}}\,\frac{\Omega_{4\pi}}{\xi R_{0.3}}.
\label{sds_R}
\eeq
is required for the cooling radius to reach $r_{\rm cool}=R$.\footnote{This value is somewhat higher than the estimate of $\dot{\Sigma}_{\star,\,\rm crit}(r_{\rm cool}=R)\simeq300\,{\rm M_\odot}/{\rm yr\,/\,kpc^{2}}$ obtained by setting $r_{\rm cool}=R$ in equation (\ref{sdscrit}), which is only strictly applicable in the $r_{\rm cool}\gg R$ limit.}
The latter is achieved and exceeded in local ULIRGs (e.g., \citealt{condon91,downes_solomon,barcos_munoz,scoville_arp220}).  As a result, radiatively cooling hot winds may be a natural explanation for the fast line emission seen in these and similar systems with very high $\dot{\Sigma}_\star$ \citep{soto_fastlines,martin_lya}. More generally, much of the star formation at high redshift occurs in systems that meet or exceed the value of $\dot{\Sigma}_{\star,\,\rm crit}$ given in equation (\ref{sdscrit}), and we therefore expect the radiative cooling of their winds to be important.

The fact that the minimum value of the mass loading required for radiative cooling to happen anywhere in the supersonic flow ($\beta_{\rm min}$) and the maximum value such that $r_{\rm cool}\rightarrow R$ ($\beta_{\rm crit}$ in eq.~\ref{betacrit}) differ only by a factor of a few suggests that the velocity of the gas when it cools is constrained. This follows from the fact that the asymptotic velocity of the hot flow is $v\simeq10^3\,(\alpha/\beta)^{1/2}\,{\rm km/s}$. Substituting $\beta_{\rm min}$ into $v$ we find a maximum velocity of the cooling flow of
\beq
v_{\rm max}(r_{\rm cool})\simeq1250\,{\rm km/s}\,\left(\frac{\alpha\,\xi\,\dot{M}_{\star,10}}{\Omega_{4\pi}\,R_{0.3}}\right)^{0.180},
\label{vmax}
\eeq
which, for extreme ULIRG-like parameters (e.g., $R=50$\,pc and $\dot{M}_\star=100$\,M$_\odot$ yr$^{-1}$; see, e.g., \citealt{barcos_munoz}) reaches $\simeq2600$\,km/s, while for LBG-like parameters ($R=1$\,kpc and $\dot{M}_\star=30$\,M$_\odot$ yr$^{-1}$) is $\simeq1200$\,km/s. Similarly, substituting $\beta_{\rm crit}$ from equation (\ref{betacrit}) into $v_\infty$ we find a characteristic velocity for cooling on scale $R$:
\beq
v_{\rm crit}(r_{\rm cool}=R)\simeq720\,{\rm km/s}\,\left(\frac{\alpha\,\xi\,\dot{M}_{\star,10}}{\Omega_{4\pi}\,R_{0.3}}\right)^{0.135},
\label{vmin}
\eeq
which reaches $\simeq1250$\,km/s and $\simeq700$\,km/s for ULIRG and LBG parameters, respectively. Although subject to the uncertainty and potential diversity in both $\alpha$ and $\Omega$ in the above equations, these scalings imply that velocities inferred from absorption and emission lines in galactic winds should have a well-defined maximum. They also imply that if one selects galaxies of approximately the same size, then $v$ should correlate with $\dot{M}_\star$ weakly. We return to the issue of velocity profiles in radiatively cooling winds in Section \ref{section:numerical}, where we show that the flow can decelerate significantly on $\sim1-100$\,kpc scales because of gravity, and in Section \ref{section:discussion}, where we argue that because of potentially super-critical mass-loading (i.e., $\beta>\beta_{\rm crit}$) fountain flows with a broad range of velocities may develop. Even so, $v_{\rm max}$ (eq.~\ref{vmax}) is a well-defined upper bound to the velocity of cooled gas in rapidly cooling hot winds that can be tested with observations, and $v_{\rm crit}(\beta=\beta_{\rm crit})$ (eqs.~\ref{vmax} and \ref{betacrit}) is a characteristic velocity for hot winds (or portions of hot winds) that are sufficiently mass loaded to cool at $r_{\rm cool}=R$ and then escape to large scales.

Both the scale of $r_{\rm cool}$ and the value of $\beta_{\rm crit}$ have additional direct consequences for observables.  If the flow becomes radiative, then it can give up at most its total thermal energy content, which varies strongly as a function of radius.  At $r=R$,  the Mach number ${\cal M}=1$ and the thermal and kinetic energy are comparable, with the total enthalpy flux exceeding the kinetic energy flux by a factor of 3.  If $\beta_{\rm crit}(r_{\rm cool}=R)$ obtains, we would therefore expect the total radiated luminosity as the matter cools to be approximately 
\beq
\eta_{\rm max}\sim\dot{E}_{\rm hot}/L_\star\sim10^{-2}, 
\label{etamax}
\eeq
where 
\beq
L_{\star}\simeq10^{11}\,L_\odot\,{\dot{M}_{\star,10}}
\label{lstar}
\eeq 
is the bolometric luminosity from star formation. This value of $\eta$ is consistent with observations of line emission seen in \cite{martin_lya} (e.g., their Figs.~14 \& 15). However, for $r_{\rm cool}\gg R$, the radiative efficiency of the cooling flow decreases rapidly because adiabatic cooling drives down the thermal content of the matter as the radius increases.  In the limit $r_{\rm cool}\gg R$, the Mach number can be approximated (for $\gamma=5/3$) as ${\cal M}^2\simeq16^{2/3}(r/R)^{4/3}$, and the radiative efficiency of the flow can be approximated as 
\beqa
\eta&=&\left.\frac{1}{{\cal M}^2}\right|_{r_{\rm cool}}=\frac{L_{X}}{\dot{E}_{\rm hot}}\simeq100 \frac{L_{X}}{L_{\star}} \nonumber \\
&\sim&5.0\times10^{-3}\,\frac{\beta^{3.89}}{\alpha^{2.84}} \,\left(\frac{{\xi\,\dot{M}_{\star,10}}}{\Omega_{4\pi}\,R_{0.3}}\right)^{1.05},
\label{eta}
\eeqa 
where $L_X$ is the luminosity radiated from the wind as it cools. The subscript $X$ represents the fact that $r_{\rm cool}$ occurs at a temperature between $10^{6.5}$ and $10^5$\,K and is thus in the X-ray or far-UV regime. In Section \ref{section:coolwind} we present a more detailed discussion of the radiative efficiency in the context of the models computed in Section \ref{section:numerical}. Figure \ref{figure:dl} shows that the radiative power of the wind is not in fact dominated by the cooling region, but by the hotter, effectively adiabatic region at smaller radius (see also eq.~\ref{lum}). Even so, equation (\ref{eta}) gives a first estimate of how the radiative efficiency of the cooling material scales with the parameters of the problem. Note, though, that this estimate for the radiative efficiency is just the thermal energy of the flow at $r_{\rm cool}$.  Once the temperature drops to $\sim10^3-10^4$\,K, it will be subject to phoionization heating from the UV emission from the galaxy or the metagalactic UV background. The material will then expand at approximately constant temperature, in photoionization equilibrium, and it will continue to radiate (this is the outer zone discussed in \citealt{silich2003,silich2004}).  We argue  in Section \ref{section:halo} that this cool outflowing material may form the cool gas seen in galactic halos at all redshifts.

The strong dependencies of $r_{\rm cool}$ on the parameters of the system translate into even stronger dependencies for the gas density, column density ($N=nr$), and emission measure (${\rm EM}=n^2r$) at the cooling radius:
\beq
n_{\rm cool}\simeq2.0\times10^{-3}\,\,{\rm cm^{-3}}\,\,
\frac{\beta^{7.34}}{\alpha^{4.76}}\frac{\xi^{1.58}}{R_{0.3}^{3.58}}\left(\frac{{\dot{M}_{\star,10}}}{\Omega_{4\pi}}\right)^{2.58},
\label{ncool}
\eeq
\beq
N_{\rm cool}\simeq2.5\times10^{19}\,\,{\rm cm^{-2}}\,\,
\frac{\beta^{4.42}}{\alpha^{2.63}}\xi^{0.789}\left(\frac{{\dot{M}_{\star,10}}}{\Omega_{4\pi}R_{0.3}}\right)^{1.79},
\label{Ncool}
\eeq
\beq
\frac{{\rm EM}_{\rm cool}}{\rm pc\,\,cm^{-6}}\simeq1.6\times10^{-2}\,
\frac{\beta^{11.8}}{\alpha^{7.39}}\,\frac{\xi^{2.37}}{R_{0.3}^{5.37}}\left(\frac{\dot{M}_{\star,10}}{\Omega_{4\pi}}\right)^{4.37}.
\label{emcool}
\eeq
The gas density at the cooling radius is important for assessing the applicability of CIE, the propensity of the medium to cool below its temperature in photoionization equilibrium, and in determining the final density of clouds precipitated out of the hot flow by thermal instability (see Section \ref{section:ti}).  The column density is important for assessing the effective absorption optical depth of a given line for absorption-line studies of winds viewed either ``down the barrel" or at large impact parameter with background quasars or galaxies. The column is also important for estimating the scattering optical depth of Ly$\alpha$ radiation and other resonant transitions in escaping the expanding flow, and for studies of emission-line halos and their velocity profiles.  Finally, the emission measure is useful for characterizing the expected surface brightness of the radiating gas in resolved systems and for emission line studies of warm/hot gas as it cools from $\sim10^{6.5}$\,K to $\sim10^3-10^4$\,K.

Note that the scalings above imply strong diversity among the observed properties of systems; even a slight change in $\beta$, $\alpha$, $\Omega$, $R$, or ${\dot{M}_\star}$ can lead to significant changes in $n_{\rm cool}$, $N_{\rm cool}$, and ${\rm EM}_{\rm cool}$.  Conversely, both the range of $\beta$ and the range of velocities for a cooling flow are tightly bounded. This narrow range of $\beta$ will imprint itself on the column density and emission measures inferred from observations. For example, substituting $\beta_{\rm min}$ into $N_{\rm cool}$, one finds a minimum column density at which cooling can occur of 
\beq
N_{\rm cool,\,min}\simeq3.4\times10^{18}\,\,{\rm cm^{-2}}\,\left(\frac{\alpha\,\dot{M}_{\star,10}}{\Omega_{4\pi}\,R_{0.3}}\right)^{0.18}\frac{1}{\xi^{0.82}}. 
\label{ncoolmin}
\eeq
If the limiting total gas column density one could detect in a given survey is $10^{20}$\,cm$^{-2}$, for example, this will significantly limit the mass loading of the systems that could in principle be seen and the velocities that material would be expected to reach. If galactic winds are better described by a wide range of $\beta$ along different sightlines in a given system, as we argue may be the case in Section \ref{section:coolwind}, then the {\it fastest} cool photoionized gas (eq.~\ref{vmax}) in a given observed system with have the {\it minimum} column density, and it will be given by equation (\ref{ncoolmin}). Indeed, the strong diversity implied by our expressions for $n_{\rm cool}$, $N_{\rm cool}$, and ${\rm EM}_{\rm cool}$ may be partially mitigated if there is a self-regulating mechanism that drives $\beta\rightarrow\beta_{\rm crit}$, as one might expect if cool clouds from the host are incorporated into the hot flow, as we discuss in Section \ref{section:rebirth}.

Finally, note that although we focus here on radiative cooling on scales larger than the host galaxy size $R$, radiative cooling can also occur at $r<R$ where the mass and energy are injected (eqs.~\ref{mdothot} and \ref{edothot}; Appendix \ref{appendix:cc85}). \cite{tenorio07} have studied this problem both analytically and numerically in the context of individual super star clusters. Assuming uniform volumetric mass loading ($\dot{m}_{\rm hot}$; g/s/${\rm cm}^3$) and energy injection rates ($\dot{e}_{\rm hot}$; ergs/s/${\rm cm}^3$), and for a fixed value of $\dot{e}_{\rm hot}/\dot{m}_{\rm hot}$, they find a critical stagnation radius $R_{\rm st}$ that approaches $r=0$ and $R$ in the limit of low and high $\dot{E}_{\rm hot}$, respectively. Only material injected with $R_{\rm st}<r<R$ participates in a supersonic outflow. Material injected at $0<r<R_{\rm st}$ rapidly cools, does not enter the outflow, and perhaps gives rise to a new generation of star formation \citep{tenorio07,palous}. As $R_{\rm st}$ approaches $R$, the total amount of material ejected in the hot outflow decreases, but the specific energy of the wind remains fixed because the ratio $\dot{e}_{\rm hot}/\dot{m}_{\rm hot}$ is fixed everywhere within the injection volume. 

In the models considered here, radiative cooling within the host galaxy will decrease the total amount of matter that participates in the outflow, decreasing the gas density, and thus increasing the critical mass loading rate needed for cooling on large scales. However, the magnitude of this effect depends on the energy and mas injection profile. For example, in CC85 and \cite{tenorio07} for star clusters, $\dot{e}_{\rm hot}$ and $\dot{m}_{\rm hot}$ are assumed to be constant throughout the volume, so that most of the matter and energy are injected very near $r\sim R$ ($\dot{M}_{\rm hot}\sim \dot{m}_{\rm hot} R^3$). Thus, $R_{\rm st}$ must be very near $R$ to significantly change the total mass loading of the outflow. Adopting a perhaps more realistic model for galaxy scales so that $\dot{m}_{\rm hot}\propto \dot{e}_{\rm hot} \propto r^{-2}$ ($\dot{M}_{\rm hot}(r)\propto r$;  i.e., an isothermal sphere; see \citealt{zhang14}), most of the matter and energy are still injected on scales of order $\sim R$. For more realistic disk-like geometries, or for extended mass and energy injection without a well-defined ``edge" (e.g., an $R$ beyond which $\dot{e}_{\rm hot}=\dot{m}_{\rm hot}=0$), more work is required to understand both the sonic point and the mass loading.

Note that for the CC85 model and the cooling function adopted, the ratio $t_{\rm cool}/t_{\rm adv}(r<R)$ is {\it largest} at $r\simeq R$. Thus our estimate of $\beta_{\rm crit}(r=R)$ in equation (\ref{betacrit}) gives the limit such that cooling is important throughout the profile, from $r=0$ outwards. This is then the limit that $R_{\rm st}\rightarrow R$ and we expect the model to break down completely because for $\beta>\beta_{\rm crit}$ the wind is radiative at all radii.

\subsection{Conduction}
\label{section:conduction}

As an aside, we note that conduction may strongly affect the temperature profiles of low-$\beta$ outflows. In the energy equation, the net sources and sinks include both radiative losses and conduction. As we have shown above, radiative losses are important for $\beta\gtrsim1$ (eqs.~\ref{betamin}, \ref{betacrit}). The net local energy deposition rate from conduction is given by 
\beq
\dot{\varepsilon}_{\rm cond}=\nabla\cdot[\kappa(T)\nabla T],
\label{cond}
\eeq
where $\kappa(T)\simeq5\times10^{-7}T^{5/2}$\,ergs cm$^{-1}$ K$^{-1}$ is the Spitzer conduction coefficient. 

Setting aside the important uncertainty concerning the strength and orientation of the magnetic field and assuming  the temperature profile follows $T\propto r^{-4/3}$ as appropriate for the CC85 model on scales $r>R$, equation (\ref{cond}) can be written as $\dot{\varepsilon}_{\rm cond}=44\kappa(T)T/9r^2$. The conduction timescale ($nk_B T/\dot{\varepsilon}_{\rm cond}$) is then
\beq
t_{\rm cond}
\simeq1.3\times10^6\,{\rm yr}\,\,\,\frac{\dot{M}_{\star,10}}{\Omega_{4\pi}}\,\frac{\beta^4}{\alpha^3}
\left(\frac{r}{R}\right)^{10/3}.
\eeq
The strong $\beta$, $\alpha$, and $r$ dependences of this expression stem directly from the normalization of the temperature in an energy driven model ($T\propto \alpha/\beta$), the radial dependence of the adiabatic temperature gradient, and the strong scaling of the conductive flux with temperature ($\propto T^{7/2}$).

The conduction timescale $t_{\rm cond}$ has a much steeper radial dependence than the advection timescale $t_{\rm adv}=r/v$. We thus expect conduction to dominate at small radii.  Setting $t_{\rm cond}=t_{\rm adv}$ (eq.~\ref{tadv}), we derive a critical radius beyond which the flow is adiabatic, and below which the flow is dominated by conduction:
\beq
r_{\rm cond}\simeq160\,{\rm pc}\,\,R_{0.3}^{10/7}\left(\frac{\Omega_{4\pi}}{\dot{M}_{\star,10}}\right)^{3/7}\frac{\alpha^{15/14}}{\beta^{3/2}}.
\label{rcond}
\eeq
The fact that $r_{\rm cond}\simeq R$ in this expression implies that $\beta\sim1$ is the critical value of the mass loading parameter below which conduction begins to dominate the temperature gradient at $R$. Setting $r_{\rm cond}=R$, we find that\footnote{This estimate ignores the strong temperature gradient in the CC85 model at $R$, which is artificially steep --- and would therefore further enhance the importance of conduction at $R$ --- because of the assumed sudden and artificial cutoff in the sources of energy and mass injection at $R$ in the CC85 model.}
\beq
\beta_{\rm crit,\,cond}\simeq0.6\,\alpha^{5/7}\left(\frac{\Omega_{4\pi}\,R_{0.3}}{\dot{M}_{\star,10}}\right)^{2/7}.
\label{betacond}
\eeq
That is, for $\beta<\beta_{\rm crit,\,cond}$, $t_{\rm cond}<t_{\rm adv}$ on scale $R$ and we expect conduction to dominate advection out to a scale given by $r_{\rm cond}$. For $\beta>\beta_{\rm crit,\,cond}$, we expect conduction to have a minor effect on the resulting solutions.  Comparing equation (\ref{betacond}) with equation (\ref{betamin}) we see that over virtually any range of $\beta$ the standard CC85 model is invalid: for $\beta\gtrsim\beta_{\rm min}$ the flow is radiative, while for $\beta\lesssim\beta_{\rm cond}$ it is highly conductive, and $\beta_{\rm min}\sim\beta_{\rm cond}$.

A full exploration of the importance of conduction for low-$\beta$ outflows is left for a future work, but here we note that a primary effect will be to flatten the temperature profile for $r\lesssim r_{\rm cond}$. In this regime we can ignore advection and radiative cooling, and thus the energy equation is integrated trivially to yield the power-law relation \citep{parker64}
\beq
T(r)\propto r^{-2/7}\,\,\,\,\,(R\lesssim r\lesssim r_{\rm cond}),
\label{parker}
\eeq
which is significantly flatter than the $r^{-4/3}$ expected for an adiabatic flow. This may have significant consequences for observations attempting to diagnose the dynamical significance of the hot flow with X-rays. For example, \cite{strickland_m82} find a flat temperature gradient potentially consistent with equation (\ref{parker}) in the M82 superwind on kpc scales. This may be due to conduction given the low value of $\beta$ implied by the hard X-ray observations of \cite{strickland_heckman} (they find $\beta\sim0.2-0.5$).

A caveat to our discussion of conduction is that the form of equation (\ref{cond}) is only valid if (1) the collisional mean free path of the electrons is smaller than the characteristic scale of the temperature gradient $\lambda_e<dr/d\ln T$, (2) the electron-proton equilibration timescale is less than the advection time, and (3) the electron thermal speed ($v_e$) is larger than the flow speed $v$. Our estimates suggest that these are all satisfied for the $\beta\gtrsim1$ models we focus on in this work, but that (1) and (2) break down rapidly for $\beta\lesssim0.2-0.5$. Thus for the minimal mass loading expected from supernova driven winds composed only of supernova ejecta ($\beta\sim0.2$) we expect the basic hydrodynamic assumption of CC85 to be invalid and a complete revaluation of the physics is in order. 
 
\begin{figure*}
\centerline{\includegraphics[width=7.3cm]{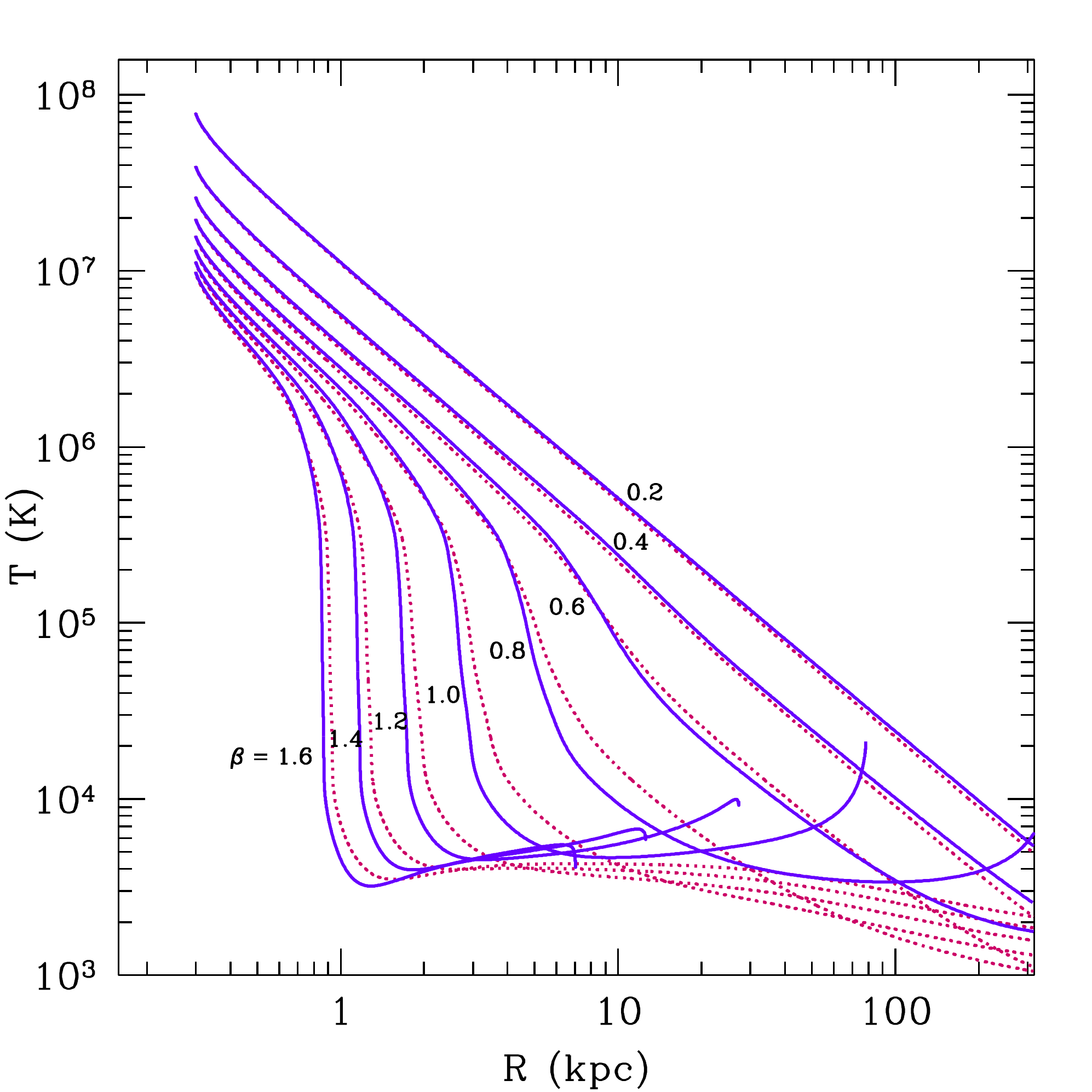}\includegraphics[width=7.3cm]{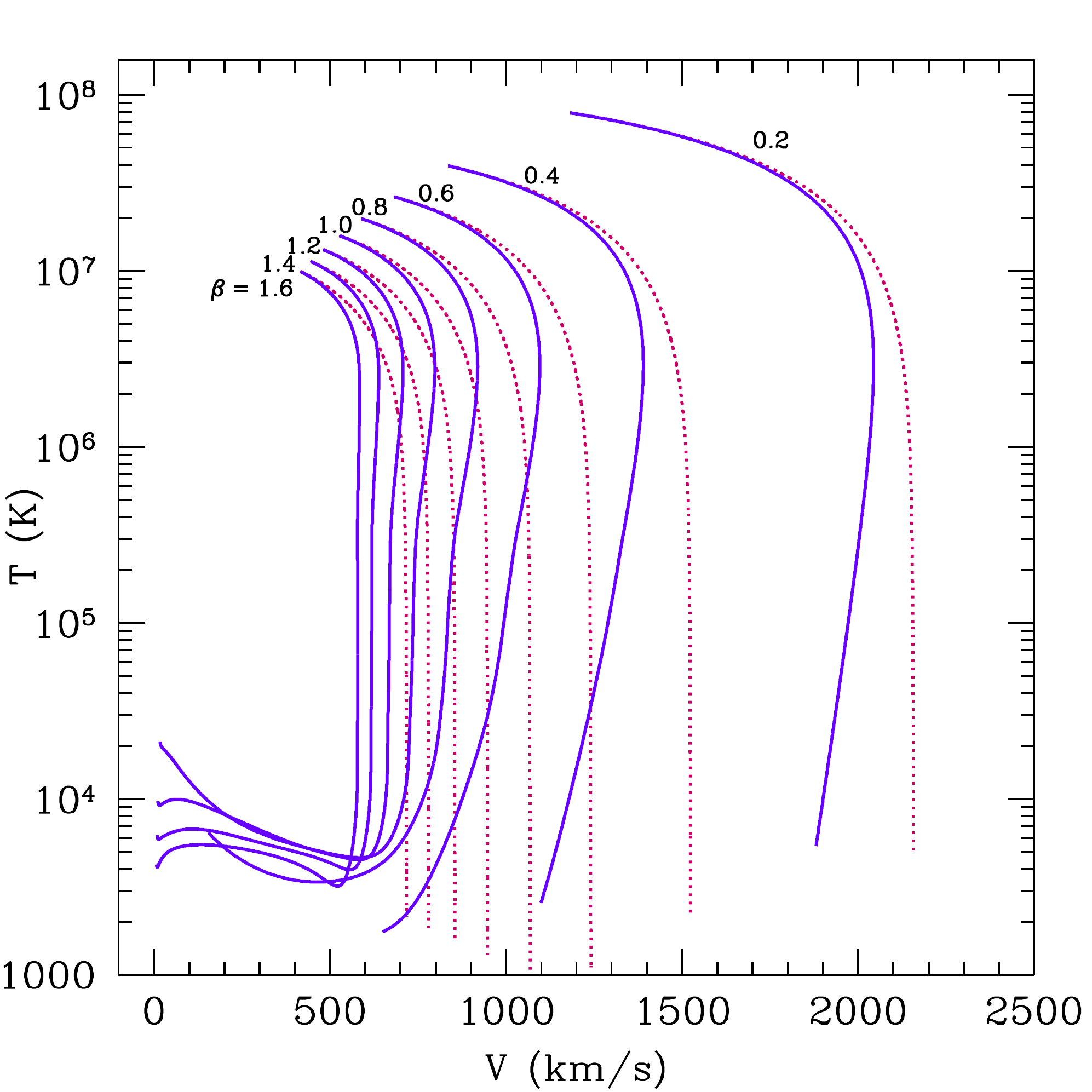}}
\centerline{\includegraphics[width=7.3cm]{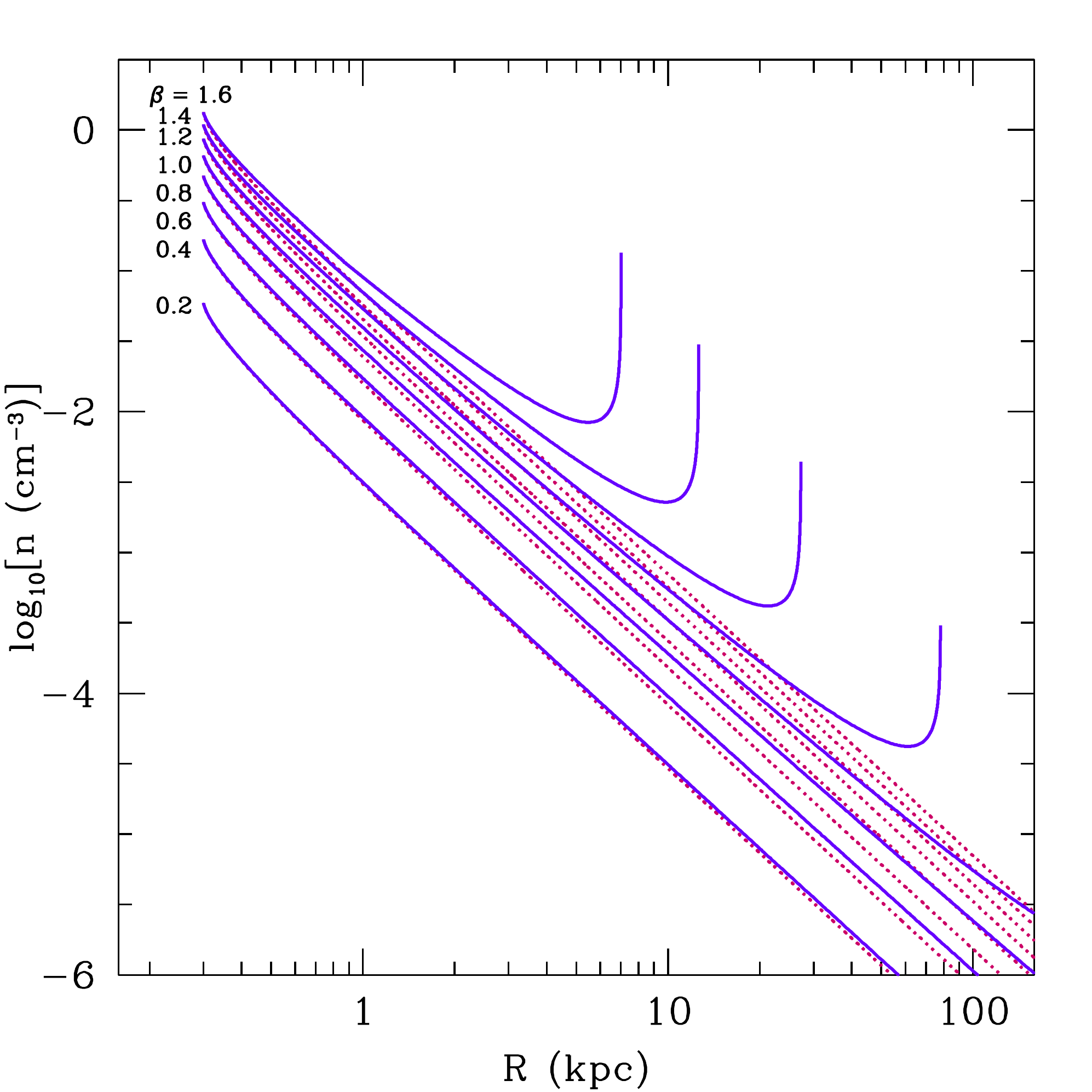}\includegraphics[width=7.3cm]{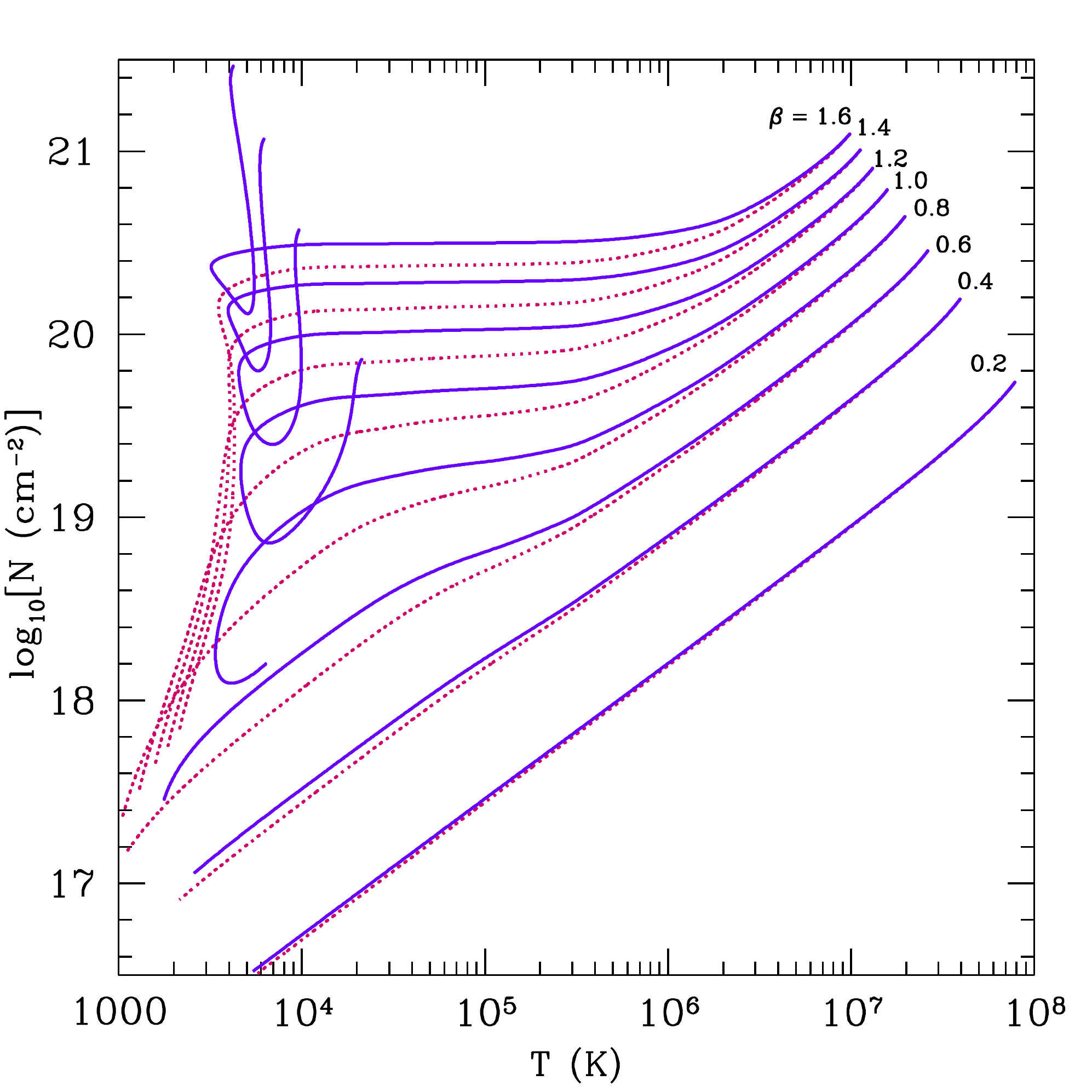}}
\centerline{\includegraphics[width=7.3cm]{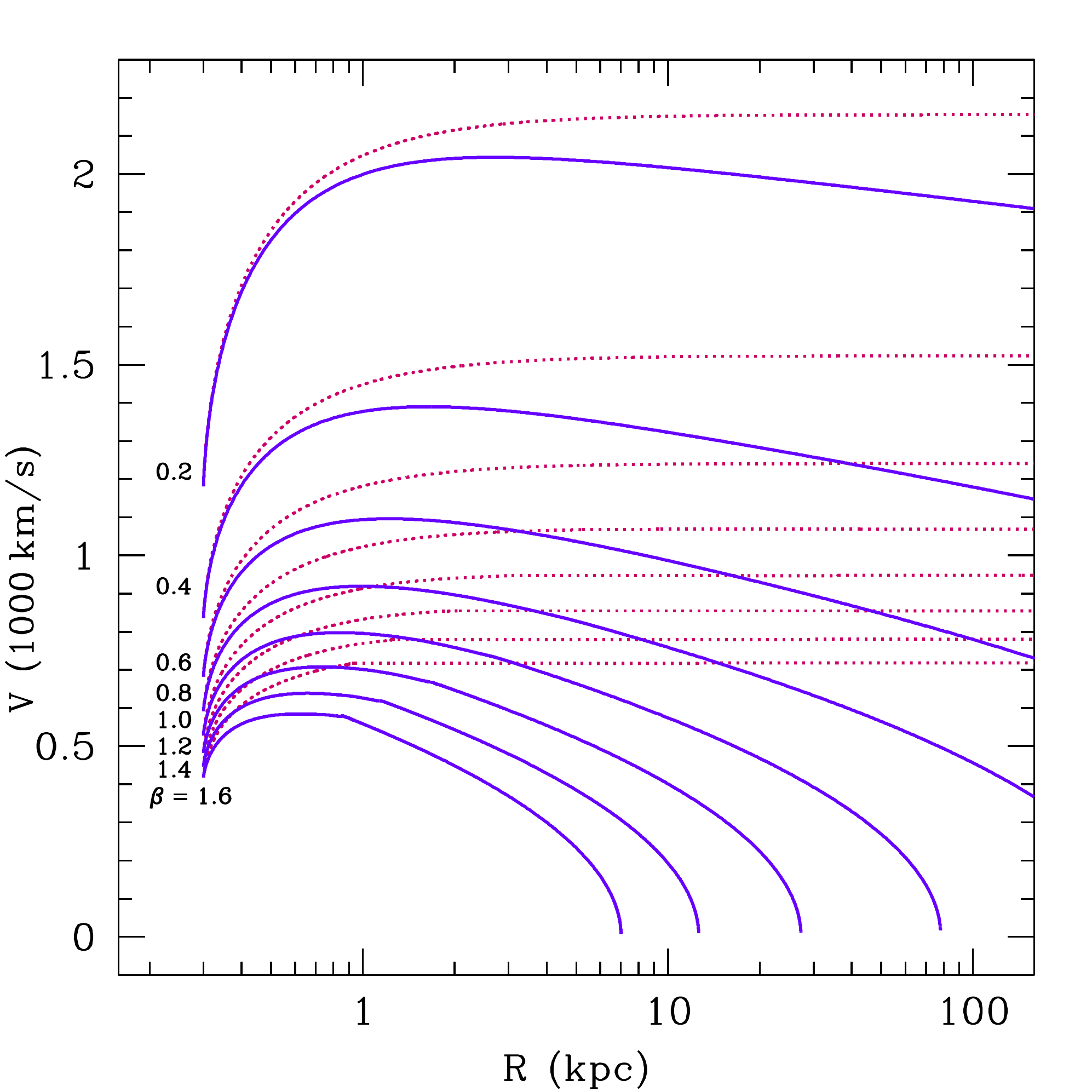}\includegraphics[width=7.3cm]{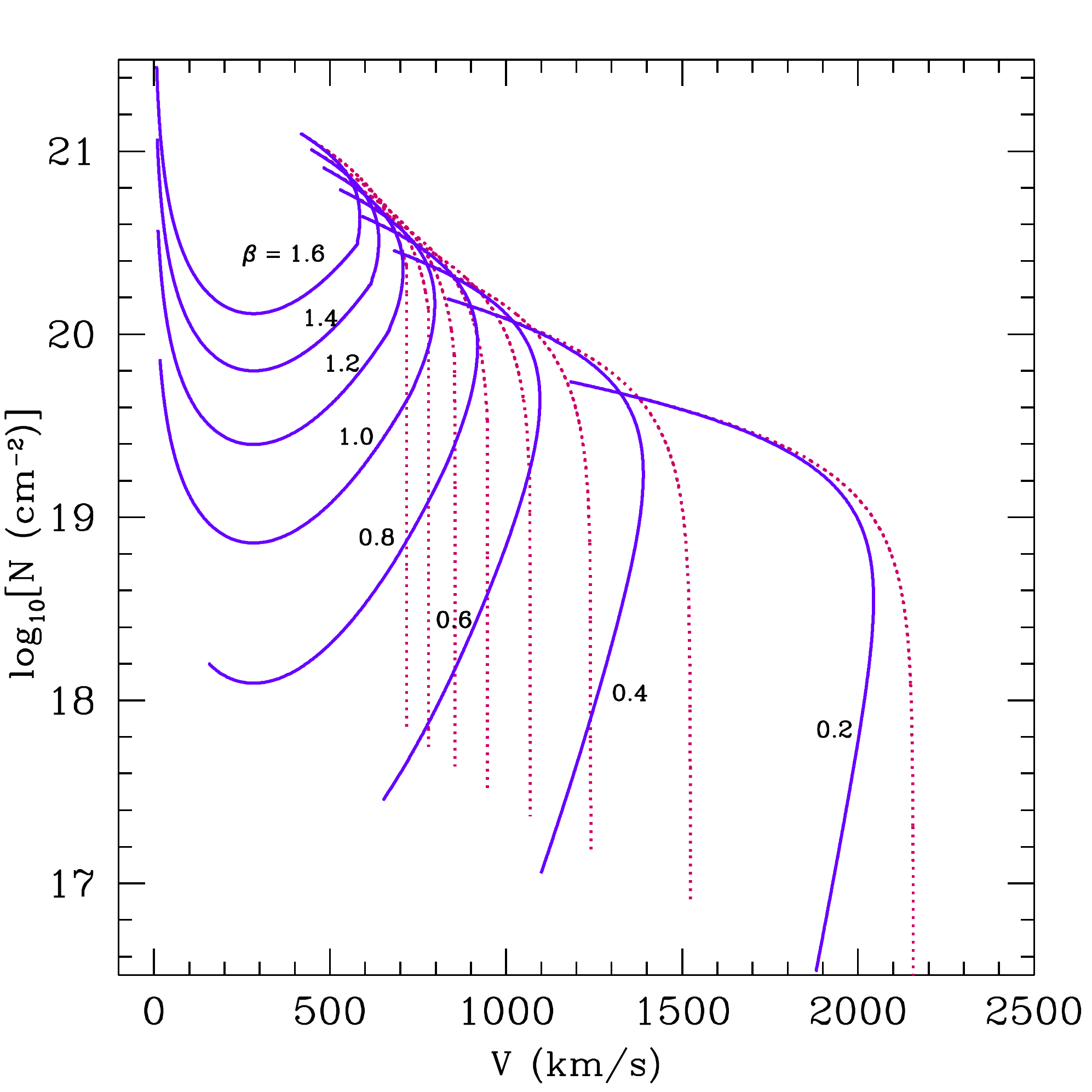}}
\caption{{\it Left Column:} Temperature (top), density (middle), and velocity (bottom) as a function of radius for a models with ${\dot{M}_\star}=10$\,M$_\odot$ yr$^{-1}$, $R=0.3$\,kpc, and $\beta=0.2-1.6$ in steps of 0.2, calculated without gravity (red dotted) and with an assumed isothermal gravitational potential with  $\sigma=200$\,km s$^{-1}$ (blue solid).  Note the rapid decrease in temperature to less than $10^4$\,K in the high $\beta$ models at the cooling radius and the decrease in velocity on large scales in the extended gravitational potential. All models assume the PIE with a meta-galactic UV background and solar metallicity \citep{oppenheimer_schaye,wiersma}. {\it Right Column:} Temperature, velocity, and column density $N=nr$ (cm$^{-2}$) versus one another for the same models. The differential luminosity as a function of temperature for these models is shown in Figure \ref{figure:dl}.}
\label{figure:t}
\end{figure*}

\subsection{A More Complete Calculation}
\label{section:numerical}

To explore the dynamics of radiative cooling in high-$\beta$ hot winds more fully, we solve the spherical steady-state wind equations with an arbitrary cooling/heating function and a simplified equation of state.  Combining the Euler equations for mass, momentum, and energy conservation, 
we find that (e.g., \citealt{lamers_cassinelli}; see Appendix \ref{appendix:wind_equations})  
\beq
\frac{dv}{dr}=\frac{v}{2r}\left(\frac{v_e^2-4c_s^2}{c_s^2-v^2}\right)+\frac{D}{C_V T}\left(\frac{\dot{q}}{c_s^2-v^2}\right)
\label{v}
\eeq
\beq
\frac{d\rho}{dr}=\frac{2\rho}{r}\left(\frac{v^2-v_e^2/4}{c_s^2-v^2}\right)-\frac{\rho}{v}\frac{D}{C_V T}\left(\frac{\dot{q}}{c_s^2-v^2}\right)
\label{rho}
\eeq
\beq
\frac{dT}{dr}=\frac{2}{r}\frac{D}{C_V}\left(\frac{v^2-v_e^2/4}{c_s^2-v^2}\right)+\frac{\dot{q}}{C_V v}\left(\frac{c_T^2-v^2}{c_s^2-v^2}\right)
\label{t}
\eeq
where $D=(T/\rho)\left. \p P/\p T\right|_\rho=k_B T/m_p$, $C_V=\left.\p E/\p T\right|_\rho=(3/2)k_B/m_p$ is the specific heat at constant volume, $c_s^2=\left. \p P/\p\rho\right|_s$ and  $c_T^2=\left. \p P/\p\rho\right|_T$ are the adiabatic and isothermal sound speeds, respectively, $\dot{q}=\dot{\varepsilon}/\rho$ is the net heating rate per gram (see Fig.~\ref{figure:c}), and $v_e=2GM(r)/r$ is the ``local" escape velocity (see Appendix \ref{appendix:wind_equations}).  For the purposes of an exploration of parameter space, we use the cooling/heating tables assuming phoionization equilibrium from \cite{oppenheimer_schaye} (see \citealt{wiersma}), which employ the \cite{haardt_madau} meta-galactic photoionizing background. In PIE, $\dot{q}$ is a function of density and temperature.  An example showing the contributions to cooling from H+He and metals separately is shown in Figure \ref{figure:c} at a characteristic density.  We neglect conductive energy transport (Section \ref{section:conduction}), which should only matter for the low-$\beta$ ostensibly adiabatic models that are not the focus of the current study. Lastly, note that the wind equations above assume that the composition of the matter does not change as a function of radius, that continuity is obeyed, and that all emission is optically-thin.

We focus on solutions that start adiabatic at $r=R$ with $t_{\rm cool}>t_{\rm adv}$.  In this limit, the Mach number of the flow is ${\cal M}(r=R)=1$, in accord with CC85 and the analysis presented in \cite{wang95a}.  We therefore solve the CC85 model for the flow at $r=R$ and then treat equations (\ref{v})-(\ref{t}) as an initial value problem and integrate to large radii using standard methods.  For the purposes of illustrating rapid radiative cooling on large scales, we assume that the wind expands into vacuum.  More discussion of a surrounding ambient medium is provided in Section \ref{section:discussion}.

Figure \ref{figure:t} shows profiles of temperature, density, and velocity for fiducial models with ${\dot{M}_\star}=10$\,M$_\odot$ yr$^{-1}$, and $R=300$\,pc, close to parameters representative of starbursts like M82 and NGC 253, or the massive star-forming clumps seen in high redshift rapidly star-forming galaxies.  The red dotted lines show the solution without a gravitational potential ($v_e(r)=0$) and the blue solid lines show the solution assuming a singular isothermal sphere ($v_e(r)=2\sigma$) with velocity dispersion $\sigma=200$\,km s$^{-1}$.  All models assume efficient thermalization $\alpha=1$, as is implied by observations of M82 \citep{strickland_heckman}, and then a range of $\beta$ from 0.2 to 1.6 in steps of 0.2.  For large $\beta$ we see the rapid decline in temperature from $\sim10^7$\,K to less than $10^4$\,K on scales of $\sim1-10$\,kpc. Here, photoionization heating balances cooling, as in the example cooling function shown in Figure \ref{figure:c}.  The material then continues to flow outward at fairly high velocity.  The right panels show that the rapidly cooling solutions have high column densities and a broad range of velocities, potentially explaining the broad emission line features seen in some systems (e.g., \citealt{soto_fastlines,martin_lya}; Section \ref{section:discussion}).

Note that the ``adiabatic" models shown with $\beta\lesssim0.6$ are meant for comparison with the more strongly radiative solutions. In fact, we expect these low-$\beta$ solutions to be strongly affected by conduction (eq.~\ref{betacond}) out to a radius given by equation (\ref{rcond}) and will thus have much flatter inner temperature profiles (eq.~\ref{parker}). For the $\beta=0.2$ model, $r_{\rm cond}\simeq1.8$\,kpc. A full solution with conduction and radiative cooling requires a complete revision of the CC85 model and is left for a future work.

Returning to the importance of radiative cooling, we see from the lower left panel of Figure \ref{figure:t} that the high-$\beta$ models in an extended gravitational potential slow down significantly on large scales \citep{wang95a}. Formally, the $\beta=1-1.6$ models spontaneously develop a sonic point at $\sim80$, 30, 10, and 7\,kpc.  This behavior follows from equation (\ref{v}) in the limit of strong cooling.  Once the temperature drops to $\sim10^4$\,K, the evolution becomes quasi-isothermal and we can approximate $\dot{q}\simeq0$  at radii $r> r_{\rm cool}$.  Taking $c_s\ll\sigma$ and $c_s\ll v$ equation (\ref{v}) shows that $dv/dr<0$ and the gas decelerates because of gravity.  The critical temperature at which the flow starts to decelerate is just the virial temperature $T\simeq3\times10^6\,\,{\rm K}\,\,\sigma_{200}^2$. Integrating from the cooling radius outwards, we derive the critical radius where the flow decelerates significantly after cooling:
\beq
r_{\rm slow}\simeq r_{\rm cool}\exp\left[\frac{3.7}{\sigma_{200}^2}\frac{\alpha}{\beta}\right],
\label{rsonic}
\eeq
where $r_{\rm cool}$ is given by equation (\ref{rcool}), and where we have used the velocity at the cooling radius in order to obtain the numerator in the exponential. We see that small changes in $\beta$, $\alpha$, or $\sigma$  change $r_{\rm slow}$ dramatically.  However, note that the detailed behavior of the gas during deceleration is sensitive to our assumption of a purely isothermal dark matter halo out to large scales.  In numerical experiments we find that if we put in a more realistic NFW-like dark matter profile, the sonic point moves out to very large radius, or does not occur within $\sim300$\,kpc for the parameters in Figure \ref{figure:t}.  Increasing $\beta$ slightly, however, can then cause strong deceleration to reappear. As is evident from equation (\ref{rsonic}) all changes in the parameters of the wind and host galaxy will strongly affect this evolution. 

The models shown in Figure \ref{figure:t} assume Solar metallicity $Z=Z_\odot$ and PIE in a \cite{haardt_madau} metagalactic background at redshift $z=0$. We experimented with changes to both. Lowering or raising the metallicity increases or decreases, respectively, the critical value of the minimum mass-loading required for strong cooling as one would expect ($\beta_{\rm min}$; eq.~\ref{betamin}). For example, for $Z=0.2\,Z_\odot$ models with $\beta=1$ and 2 are qualitatively similar to $Z=Z_\odot$ models with $\beta\simeq0.6$ and 1.4, respectively. Note that the shape of the cooling function allows for rapid radiative cooling even in primordial gas (H$+$He in Fig.~\ref{figure:c}): a model with $Z=0$ and $\beta=2$ resembles the model with $\beta=1$ shown in Figure \ref{figure:t}. Conversely, a model with $Z=4Z_\odot$ and $\beta=1$ has a cooling radius at the same location as the $\beta=1.4$ lines in Figure \ref{figure:t}. Changes to the redshift for the PIE calculation have essentially no effect on the location of the cooling radius or the critical $\beta$ for cooling, but they do change the post-cooling temperature in accordance with what one would expect for a harder metagalactic background.

Over much of parameter space the cooling radius occurs at a sufficiently high density and temperature that employing a standard CIE cooling function instead of PIE does not change the cooling radius significantly. However, the post-cooling evolution is completely different. In CIE, the gas expands adiabatically after cooling with $T\propto r^{-4/3}$, which differs qualitatively from the approximately isothermal evolution on large scales for the models shown in Figure \ref{figure:t}.

Finally, note that the flow time to $100$\,kpc is $\gtrsim100$\,Myr and we expect the assumption of steady-state conditions to be violated on this timescale, at least in starburst systems.  We further expect the wind to expand into a local circumgalactic medium, sweeping up and shocking matter in a galaxy-scale wind-blown bubble (see Section \ref{section:halo}).

\subsection{Thermal \& Convective Instability}
\label{section:ti}

\cite{bs89} (hereafter BS89) studied the case of thermal instability (TI)  in a dynamically evolving background, which is the case applicable to TI in galactic winds.    The key radial range where TI is important is where the cooling time is less than the flow time.    Otherwise there is insufficient time for the instability to grow significantly.   Under these conditions, both convection and isobaric TI can in principle grow.   The amplification of density fluctuations by convection is given by a factor \citep{bs89}
\beq
A_{\rm conv} \sim \exp\left[\int |N| dt\right] = \exp\left[\int |N| \, t_{\rm cool} \frac{dt}{t_{\rm cool}}\right]
\label{ampconv}
\eeq
where 
\beq
N^2(r) =\frac{2}{5}\frac{m_p}{k_B}\left(-\frac{1}{\rho}\frac{dP}{dr}\right)\frac{ds}{dr},
\eeq
$N(r)$ is the Brunt-V\"ais\"ala frequency, which characterizes convective instabilities, $s$ is the entropy per unit mass, and the integral over time in equation (\ref{ampconv}) is co-moving with the flow so that $dt = dr/v$.\footnote{Anisotropic thermal conduction along magnetic fields lines modifies the convective instability criterion in dilute plasma such as the galactic winds of interest here (Quataert 2008).   This does not change our  conclusion that convection is relatively unimportant in galactic winds because the timescale argument applies even to conduction-mediated convection.}   Evaluating this over the radial range where initially hot galactic winds cool appreciably, we find that $\int |N| dt$ is typically $\ll 1$.   Thus there is insufficient time for convection to grow appreciably.   This is fundamentally because the Brunt-V\"ais\"ala frequency introduced by radiative cooling in a wind is given roughly by $N t_{\rm cool} \sim \mathcal{M}^{-1}$ where $\mathcal{M}$ is the Mach number of the flow and the flow only spends a time $dt \sim t_{\rm cool}$ at radii where cooling is significant.  Note that at large radii once the flow comes into PIE and $T \sim$ constant, $ds/dr > 0$ and the flow is convectively stable.

If the cooling rate is assumed to scale as $\propto n^2 \Lambda(T)$, the amplification of density fluctuations by isobaric thermal instability is given by a factor \citep{bs89}
\beq
A_{\rm TI} \sim \exp\left[\frac{3}{5} \int \left(2 - \frac{\partial \ln \Lambda}{\partial \ln T}\right) \frac{dt}{t_{\rm cool}}\right].
\label{ampTI}
\eeq
Evaluating this for the cases shown in Figure \ref{figure:t}, we find that $A_{\rm TI} \sim 30-300$ (for $\beta=1.0-1.6$, respectively), with all of the contribution due to the small radial range where the flow cools abruptly from $T\sim 10^{6}$ K to $\sim 10^3-10^4$ K (see upper left panel of Fig.~\ref{figure:t}).  Physically, as the flow cools from high to low temperature, each logarithmic interval in temperature (which corresponds to $dt \sim t_{\rm cool}$) contributes an order unity contribution to the integral in equation (\ref{ampTI}), so that the net amplification is set by the few decades in temperature decrease during the rapid cooling of the wind.

The fact that the amplification of density fluctuations by TI is limited to a factor of $\sim 100$ implies that for the medium to develop a multi-phase structure there must be initial fluctuations of finite amplitude present in the flow.  Given the high initial sound speed in the thermally-driven wind, one might expect any such fluctuations to have been erased by the time rapid cooling sets in.  However, the hot outflow quickly becomes supersonic and thus expands outwards more quickly than it can establish pressure equilibrium.  In addition, the mixing of cool gas into the hot flow will likely imprint large inhomogeneities on the flow, which will seed both instabilities and rapid cooling (see Section \ref{section:rebirth}).  An analogous situation can be seen in Fig A3 of \cite{sharma} where an inhomogeneous cooling inflow (rather than a wind) with $t_{\rm cool} \lesssim t_{\rm adv}$ and initial density fluctuations generates significant multiphase structure.

Another requirement for equation (\ref{ampTI}) to apply is that the fluctuations must be isobaric, rather than isochoric.    This implies that there must be initial density fluctuations over a spatial scale 
\beq
L\lesssim \frac{H_{\rm cool}}{\mathcal{M}}
\label{size}
\eeq
where $H_{\rm cool}$ is the width of the radial region where cooling is rapid (the temperature scale height) and $\mathcal{M}$ here is for the hot wind prior to the onset of very rapid cooling.    Assuming this is satisfied, TI can in principle produce final densities of 
\beq
n_{f} \sim 100 \, n_{\rm cool} \, \left(\frac{10^4 \, {\rm K}}{T_f}\right) \left(\frac{T_{\rm cool}}{10^6 \, {\rm K}}\right)
\label{nti}
\eeq 
where $n_{\rm cool}$ and $T_{\rm cool}$ are the density and temperature of the hot flow at the cooling radius $r_{\rm cool}$ (see eq.~\ref{ncool}).  However, because all of the gas ultimately cools to PIE in our fiducial models at $\sim 10^3-10^4$\,K there is nothing to maintain pressure equilibrium for gas having $n_f \gg n_{\rm cool}$.   Gas at different density will have slightly different temperatures in PIE, which will allow some density fluctuations even if pressure equilibrium is satisfied. More importantly,  some of the cold denser gas produced by TI may not have time to expand significantly on the flow time.   One way to see this is to note that if  clouds occupy a fraction of $10^{-2}$ of the area of the sphere but have densities of $\sim 100 n_{\rm cool}$, they will carry the same mass flux as the original hot wind.   With that  filling fraction and a post cooling temperature of $10^3-10^4$\,K the gas will not have time to expand to fill the sphere on the flow time.   The proper physical picture may thus be one in which the cool clouds produced by TI in galactic winds are initially over-pressurized relative to their surroundings and then expand out at $\sim10$s of km/s, eventually re-establishing pressure balance.  Alternatively, magnetic and/or cosmic-ray pressure may be important in confining the cool gas.

\section{Discussion}
\label{section:discussion}

Two interconnected puzzles persist in the physics of cool gas in galactic winds and halos that may be solved by rapidly cooling, thermally-unstable hot winds of the type presented in Figure \ref{figure:t} and Section \ref{section:winds}.  The first is  the presence of fast, but cool, outflows observed in emission and absorption in winds from rapidly star-forming galaxies across the universe. How is the molecular, neutral atomic, and ionized gas accelerated to the hundreds and even several thousands of km/s in some systems without being shocked, shredded, and incorporated into the hot flow? The second is the pervasive warm neutral atomic and ionized gas seen in the halos of galaxies at low and high redshift \citep{steidel_cgm,werk}. What is the origin of this gas, and how does it persist on large scales without rapidly accreting?

\subsection{Cool Gas in Winds}
\label{section:coolwind}

The prevailing picture of the cool gas in winds is that it is entrained from the host ISM in the hot wind, and then accelerated from the galaxy by ram pressure (e.g., \citealt{veilleux_review}).  In accord with this view, the soft extraplanar X-ray emission from wind galaxies is often interpreted as shocked interfaces between the cool clouds and the surrounding hot medium (e.g., \citealt{strickland_ngc253a,strickland_ngc253b}). However, it is unclear how the cool gas clouds are accelerated to the velocities and large scales seen without being shredded by hydrodynamical instabilities and incorporated into the hot flow \citep{klein,cooper09,marcolini,mccourt,scannapieco,schneider}.  

Figure \ref{figure:t} suggests an alternative interpretation of both the soft X-ray emission and the cool gas seen in galactic winds. In this picture, a hot flow with $\beta\gtrsim1$ expands from the energy injection region and cools radiatively on large scales \citep{wang95a}.  We propose that the high velocity cool clouds seen in absorption line tracers like NaD, Mg, and other optical and UV resonance lines are precipitated directly from the hot gas in a rapidly cooling flow. Furthermore, the observed extraplanar soft X-ray emission may also originate from cooling, and may not necessarily be dominated by interfaces and shocks as in the standard picture. 

Viewed ``down the barrel" towards the central star-forming host, one would expect to see fast, predominantly blue-shifted absorption lines of partially ionized gas and soft X-ray emission. Emission lines are also possible (see eq.~\ref{emcool}). Indeed, recently \cite{martin_lya} suggested that the very fast Ly$\alpha$ emission seen from local ULIRGs could originate from direct radiative cooling of the hot gas in a CC85-like outflow \citep{soto_martin_diagnostic,soto_fastlines}. We find a critical surface density of star formation in the ULIRG range --- $\dot{\Sigma}_\star\simeq500$\,M$_\odot$ yr$^{-1}$ kpc$^{-2}$ --- for rapid cooling on a scale of a few hundred parsecs, given high thermalization efficiency (eq.~\ref{sds_R}). A particularly interesting prediction of the model is a direct connection between the velocity of the material and its mass loading. Higher $\beta$ implies more rapid cooling closer to the source, but also smaller velocity since $v\propto\,(\alpha/\beta)^{1/2}$ in the CC85 model (see eqs.~\ref{vmax} and \ref{vmin}). For ULIRGs like those discussed in \cite{soto_fastlines} and \cite{martin_lya}, these scalings imply rapid cooling and bright line emission at $\gtrsim 10^3$\,km/s, as is observed. Although speculative, we note that for ULIRG-like parameters and with $\beta\rightarrow\beta_{\rm crit}$ (eq.~\ref{betacrit}), $n_{\rm cool}$ and $N_{\rm cool}$ (eqs.~\ref{ncool} and \ref{Ncool}) reach $\sim10^{2.5}$\,cm$^{-3}$ and $10^{22.5}$\,cm$^{-2}$, respectively, implying that after thermal instability the material may self-shield and rapidly form molecules ($n_f\sim10^2n_{\rm cool}\sim10^{4.5}$\,cm$^{-3}$; eq.~\ref{nti}), depending on its dust content. The possibility that molecules may form directly in a rapidly cooling hot flow should be assessed by following grain sputtering, grain-gas cooling/heating, and line cooling below $10^4$\,K. Much less speculative is the fast atomic line emission and absorption with maximum velocity given by equation (\ref{vmax}) that should be seen from rapidly star-forming systems if the emission measures and column densities are high enough, respectively. This picture of fast cool outflows may provide an explanation for the high velocities measured for objects like those from \cite{tremonti,diamond,sell,geach} and should be tested with absorption line observations from a large variety of systems \citep{ham}.

\begin{figure}
\centerline{\includegraphics[width=8.5cm]{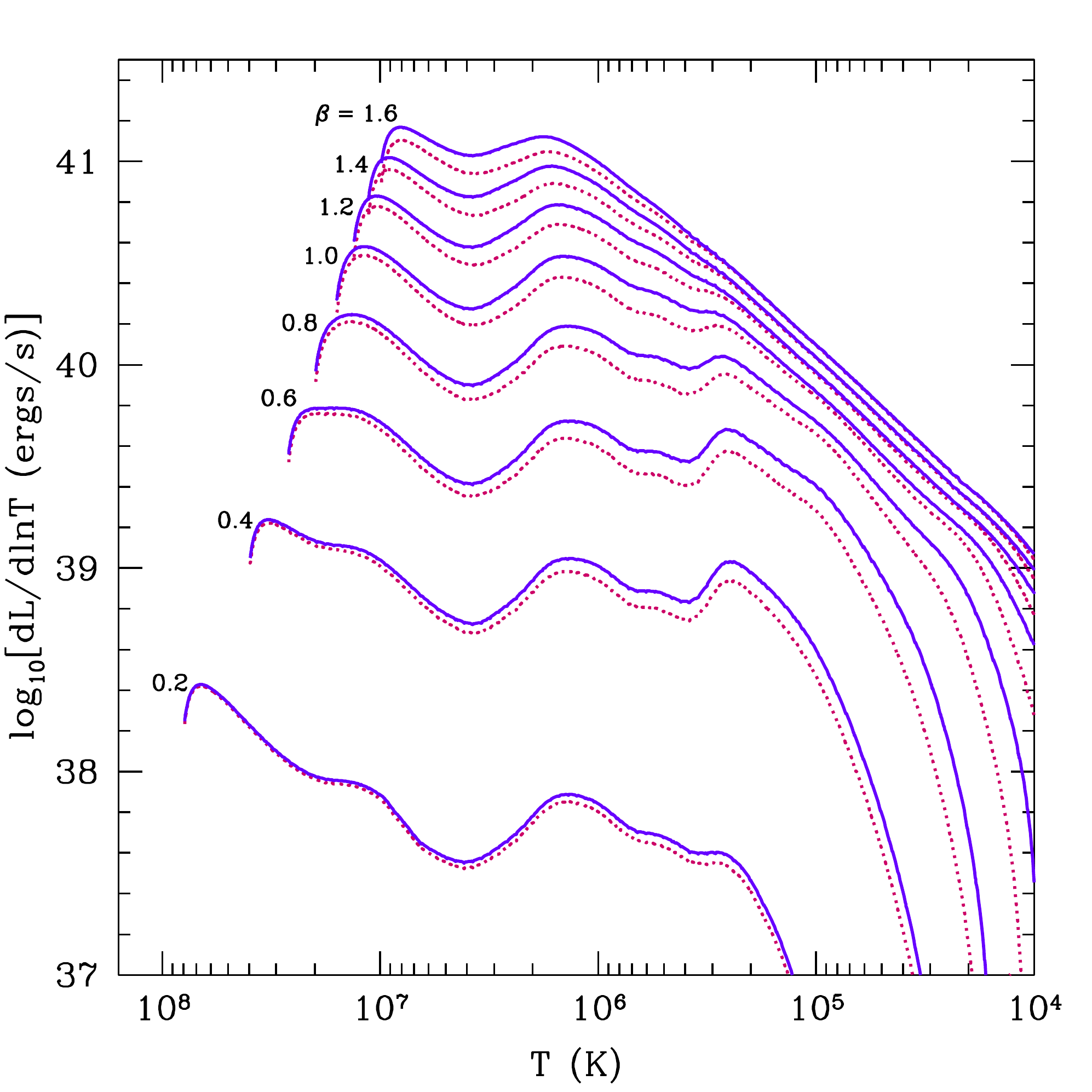}}
\caption{$dL/d\ln T$ as a function of temperature for the wind models shown in Figure \ref{figure:t}. The power-law form $dL/d\ln T\propto T$ given by equation (\ref{lum}) in the limit of strong cooling is evident for the high-$\beta$ solutions. Note that these models have $\dot{M}_\star=10$\,M$_\odot$ yr$^{-1}$, which implies a bolometric luminosity of $L_{\rm bol}\sim10^{44.5}$\,ergs s$^{-1}$ and a corresponding energy injection rate of $\dot{E}_{\rm hot}\sim10^{42.5}$\,ergs s$^{-1}$, both of which are substantially off the top of the plot. Most of the injected energy goes to adiabatic losses rather than radiative cooling.}
\label{figure:dl}
\end{figure}

An observational prediction of the radiative wind picture advocated here is that the observed X-rays arise from recombination, with a well-defined temperature progression from hot to cool, possibly with non-equilibrium ionization physics, and then a spatially extended region of photo-ionized gas together with the product of the non-linear evolution of the thermal instability in a supersonic background (Section \ref{section:ti}). Such a picture might help explain the strong spatial correspondence between the H$\alpha$ and soft X-ray emission in galactic winds \citep{strickland_xa, strickland_xb}. We highlight the fact that for a medium with constant mass flux cooling from high to low temperature, there is a particularly simple relationship between the total energy radiated per logarithmic interval in temperature and the mass flux, given by 
\beq
\frac{dL}{d\ln T}\simeq\frac{3}{2}\dot{M}_{\rm hot}\frac{k_B T}{m_p}.
\label{lum}
\eeq
This expression follows from dropping the first term on the right hand side of equation (\ref{t}) and taking the limit that the medium is highly supersonic ($v^2\gg c_s^2$, $c_T^2$).\footnote{Note that in the subsonic limit $c_T^2/c_s^2=1/\gamma=3/5$ and the numerical factor on the right hand side of equation (\ref{lum}) changes from $3/2\rightarrow5/2$ (e.g., \citealt{fabian94}).}  Figure \ref{figure:dl} shows the differential luminosity $dL/d\ln T$ along the temperature profile for the models in Figure \ref{figure:t}. The expected power-law $dL/d\ln T\propto T$ in equation (\ref{lum}) is recovered for the high-$\beta$ models. We have numerically verified that equation (\ref{lum}) is precisely satisfied in the cooling region during the precipitous drop in $T$. Thus, if the cooling region is identified in observations, equation (\ref{lum}) provides a convenient expression for estimating the mass loss rate directly from X-ray observations.

As a nearby observational example of a system that might showcase the onset of strong radiative cooling in a galactic wind, we consider the $\sim0.3-0.6$\,kpc long conical limb-brightened frustum defined in \cite{strickland_ngc253a} along the minor axis to the south of NGC 253.  Using the count rate in the energy range $0.3-1$\,keV (0.06 counts/s) or $0.3-2$\,keV (0.07 counts/s), temperature (0.5\,keV), and volume reported by \cite{strickland_ngc253a} for this region (assuming a distance of $D=2.6$\,Mpc), one derives a total radiated flux of $\sim6-10\times10^{-13}$\,ergs cm$^{-2}$ s$^{-1}$ and a hot gas density of order $\sim0.05-0.1$\,cm$^{-3}$ for gas with unity filling factor in CIE with Solar abundances. Comparing to the gas pressure and density derived from H$\alpha$ observations by \cite{mccarthy} in the same region,  \cite{strickland_ngc253a} argue that the true hot gas density is $\sim0.1-0.3$\,cm$^{-3}$. These densities imply a cooling timescale of $\sim3-9$\,Myr, whereas the advection time is of order Myr for gas at 500\,km/s. In general, we find that our radiative wind solutions have $t_{\rm cool}/t_{\rm adv}\sim3-10$ at $r\sim 1-2R$ at temperatures of $k_{\rm B}T\simeq0.5$\,keV and that the ratio decreases rapidly at larger radius because $\Lambda\propto T^{-0.7}$, as described in Section \ref{section:analytic}. The NGC 253 observations therefore appear consistent with our models in which radiative cooling becomes important at larger scales. We thus propose that the H$\alpha$ and X-ray emission from this region and beyond may be the result of the non-linear development of the thermal instability described in Section \ref{section:ti}. The fact that the H$\alpha$-emitting clouds and the hot gas are in pressure equilibrium, and that the H$\alpha$ flux increases beyond this region (see Fig.~5 of \citealt{strickland_ngc253a}, right panel) supports the notion that $r_{\rm cool}\sim0.5-1$\,kpc. The extended multi-kpc soft X-ray and line-emitting halo would then be interpreted as the large-scale aftermath of strong radiative cooling and thermal instability on sub-kpc scales. 

Note that if this feature in NGC 253 were to represent a region of strong radiative cooling, the hot gas mass loss rate implied by equation (\ref{lum}) is quite small, of order $\sim0.01-0.03$\,M$_\odot$/yr, which would imply a small value for $\beta$ if $\dot{M}_\star\sim5$\,$M_\odot$/yr and a potentially very large wind velocity if $\alpha\sim1$. However, as estimated above, $t_{\rm cool}/t_{\rm adv}\sim3-10$ for an assumed velocity of 500\,km/s, implying that equation (\ref{lum}) does not apply. The estimated $\dot{M}_{\rm hot}$ should be larger by the ratio $t_{\rm cool}/t_{\rm adv}$, which would increase the implied $\beta$ to $\sim0.1$. The additional complications made clear by the beautiful X-ray image of the frustum from \cite{strickland_xb} is first the importance of the opening angle $\Omega$ --- a naive estimate indicates $\Omega\sim0.01-0.05$ --- and second the potential importance of non-spherical areal divergence as the flow expands on $\sim0.5-{\rm few}$\,kpc scales \citep{suchkov94,suchkov96}. The latter may be especially important as the hot wind expands laterally just above the disk and encounters the halo.

Although the example of NGC 253 is illustrative, the details of what is observed will also depend on the photoionizing spectrum seen by the gas, the mass-loading along any given sightline, and whether or not the flow maintains ionization equilibrium (e.g., \citealt{breitschwerdt94,breitschwerdt99}).  Clearly much more work is required to fully assess strong radiative cooling as an explanation for the H$\alpha$ and soft X-ray halos of NGC 253, M82, and other local starbursts. In fact, \cite{strickland_ngc253b} argued against the possibility of radiative cooling for the origin of the cool gas in NGC 253, but did so using the cooling times derived on multi-kpc scales, where we would argue $r\gg r_{\rm cool}$. The story of M82 may be more complicated because the wind has likely overrun halo gas from the tidal interaction with M81, or because it simply does not fit with the model of a radiatively cooling wind discussed here. Work by \cite{hoopes_OVI_M82} indicates weak cooling on large scales. In addition, \cite{strickland_heckman} found that $\alpha\sim1$ and $\beta\sim0.2-0.5$ in M82  based on the diffuse hard X-ray emission observed in its core. Although this value of $\beta$ is nominally below our analytic requirement for strong radiative cooling on kpc scales for Solar metallicity gas, the wind may undergo distributed mass-loading and entrain material on the scale of $R$, as we discuss below in Section \ref{section:rebirth}. In this way, the small value of $\beta$ measured in the core may be compatible with strong radiative cooling (higher $\beta$) on somewhat larger scales. Much more work is required to assess the possibility of a radiative flow in the many systems for which a careful comparison is possible (e.g., VV 114, \citealt{grimes_vv114}; Mrk 231, \citealt{veilleux_mrk231,rupke_mrk231}; and many more; \citealt{ham}), but the example of NGC 253 implies that very high spatial resolution observations may be required to localize the cooling region in mass-loaded starburst winds with X-ray observations.

An additional zeroth-order observational constraint on radiative hot winds comes from the total diffuse X-ray emission observed from galaxies (as in \citealt{zhang14}). As discussed in Section \ref{section:analytic}, the rapidly cooling wind can radiate at most $\dot{E}_{\rm hot}$ power, which for steady-state star formation is $\sim1$\% of the bolometric power of star formation (eq.~\ref{etamax} and \ref{eta}), $L_\star=\epsilon\dot{M}_\star c^2$, where $\epsilon\simeq7\times10^{-4}$ is an IMF-dependent constant. In practice, the radiative efficiency is much lower than this 1\%. For example, even the $\beta=1.4$ model shown in Figure \ref{figure:t} radiates $\simeq0.05\dot{E}_{\rm hot}\simeq5\times10^{-4}\,L_\star$ during the cooling phase, whereas the $\beta=1$ model radiates $\sim5$ times less. Figure \ref{figure:dl} summarizes these results. It also shows that the $dL/d\ln T$ distribution is quite flat for low-$\beta$ models and that most of the energy is radiated between $10^{6}-10^{7.5}$\,K temperatures, where the flow is effectively adiabatic, even for high-$\beta$.

These numbers for the radiative power of cooling winds can then be compared with observations of the total (integrated) soft X-ray emission from rapidly star-forming galaxies.  For example, \cite{mineo} report the 0.5-2\,keV X-ray luminosity of star-forming galaxies to be $L_X\simeq5\times10^{38}\dot{M}_\star$\,ergs s$^{-1}$, implying $L_X/L_{\star}\simeq5\times10^{-5}$. In an attempt to approximate the bolometric X-ray luminosities of their galaxies from 0.5-10\,keV, they find a number $\sim20$ times higher, $L_X/L_{\star}\sim10^{-3}$. These numbers imply that rapid wind cooling could contribute significantly to the total X-ray budget of rapidly star-forming galaxies in extreme cases, but even the most mass-loaded model in Figure \ref{figure:t} does not violate an upper bound from observations. For fixed mass loading, system radius, and thermalization efficiency, we note that the radiative efficiency scales nearly linearly with the star formation rate (eq.~\ref{eta}). A particularly interesting case for a more careful observational comparison is NGC 6240 where the diffuse hard X-ray emission is larger \citep{wang_6240}.

\subsection{Distributed Mass Loading and Cloud Rebirth}
\label{section:rebirth}

The fact that cool clouds are readily destroyed by KH instabilities motivates a scenario for distributed mass loading and cool cloud production in hot winds. In this picture, supernova ejecta first thermalize on small scales and with a small hot gas mass loading factor of $\beta\sim0.2$ (as appropriate for the mass loss from supernovae alone) producing conditions for an initially metal-rich wind ($Z\sim$ few times Solar for supernova ejecta, and $\alpha$-element enriched). This picture is similar to that discussed in \cite{suchkov96}. The initially low-$\beta$ outflow is very hot and fast since the velocity and temperature of the wind scale as $v\simeq2000(\alpha0.2/\beta)^{1/2}$\,km/s and $T\simeq9\times10^7(\alpha0.2/\beta)$\,K, respectively. This wind sweeps up, shocks, shreds, and incorporates cool gas on the scale of the host galaxy $R$, with four important consequences: (1) $\beta$ increases, (2) $T$ decreases, (3) $v$ decreases, (4) $Z$ decreases. Note that the first three of these effects {\it decrease} the cooling timescale relative to the advection timescale on larger scales, increasing the chances for strong radiative cooling as in Figure \ref{figure:t}. In particular, the increase in $\beta$ drives the cooling radius $r_{\rm cool}$ rapidly inwards to small radii (eq.~\ref{rcool}). In addition, the swept up clouds seed the flow with density perturbations, which may grow via the thermal instability at $r_{\rm cool}$ (eq.~\ref{nti}). In essence, the rapid destruction of cool clouds on small scales initiates their subsequent formation on larger scales, an almost literal transmigration: self-induced cool cloud rebirth after death in an earlier form.

This scenario for producing cool fast gas from hot winds is complicated by the very strong dependencies outlined in Section \ref{section:winds}, and by the multi-dimensional and time-dependent character of real galactic winds.   Indeed, rather than thinking of a galaxy as having a single $\alpha$, $\beta$,  and $\Omega$, we would instead argue for a picture in which winds with a wide range of $\beta$ and $\alpha$, subtending fractions of $4\pi$, emerge from a given galaxy. Along some sightlines the amount of gas swept up is small, $\beta$ stays close to $\simeq0.2$, $\alpha$ is high, and the very fast, hot, and adiabatic wind escapes to large scales. Along other sightlines the amount of material swept up is enough to quench the wind completely: the velocity decreases sufficiently and the flow becomes sufficiently radiative ($\beta \gtrsim\beta_{\rm crit}$; eq.~\ref{betacrit}), that the outflow turns into a mass-loaded fountain on kpc scales. Thus, in looking at the upper left panel of Figure \ref{figure:t} one might imagine all $\beta$s emerging from a single system, and not just one; the observations will then be a convolution of these diverse outflows and their interactions.

This picture has interesting consequences. It suggests that the most rapidly cooling material with the strongest mass-loading, and hence the highest column densities and emission measures (eqs.~\ref{ncool}, \ref{emcool}), {\it that can in fact escape to large scales} will have velocities correlated with the local escape velocity. In this way, the process of cool gas incorporation may be self-limiting: if too much gas is swept up from the host, the wind never emerges. This process may be related to the observation by \cite{martin_2005} that the wind velocity and escape velocity are correlated. Also consistent with this picture, detailed absorption line profiles from wind galaxies in general imply a broad distribution of column densities and velocities, with the fastest material having the lowest column density and vice-versa (e.g., \citealt{rupke_2005a,rupke_2005b}; but there are notable exceptions, e.g., \citealt{diamond}). A prediction of this picture is that the limiting $\beta$ would be given in terms of observables by equation (\ref{betacrit}), with a weak dependence on host size and star formation rate at fixed $\alpha$.

\begin{figure}
\centerline{\includegraphics[width=8.5cm]{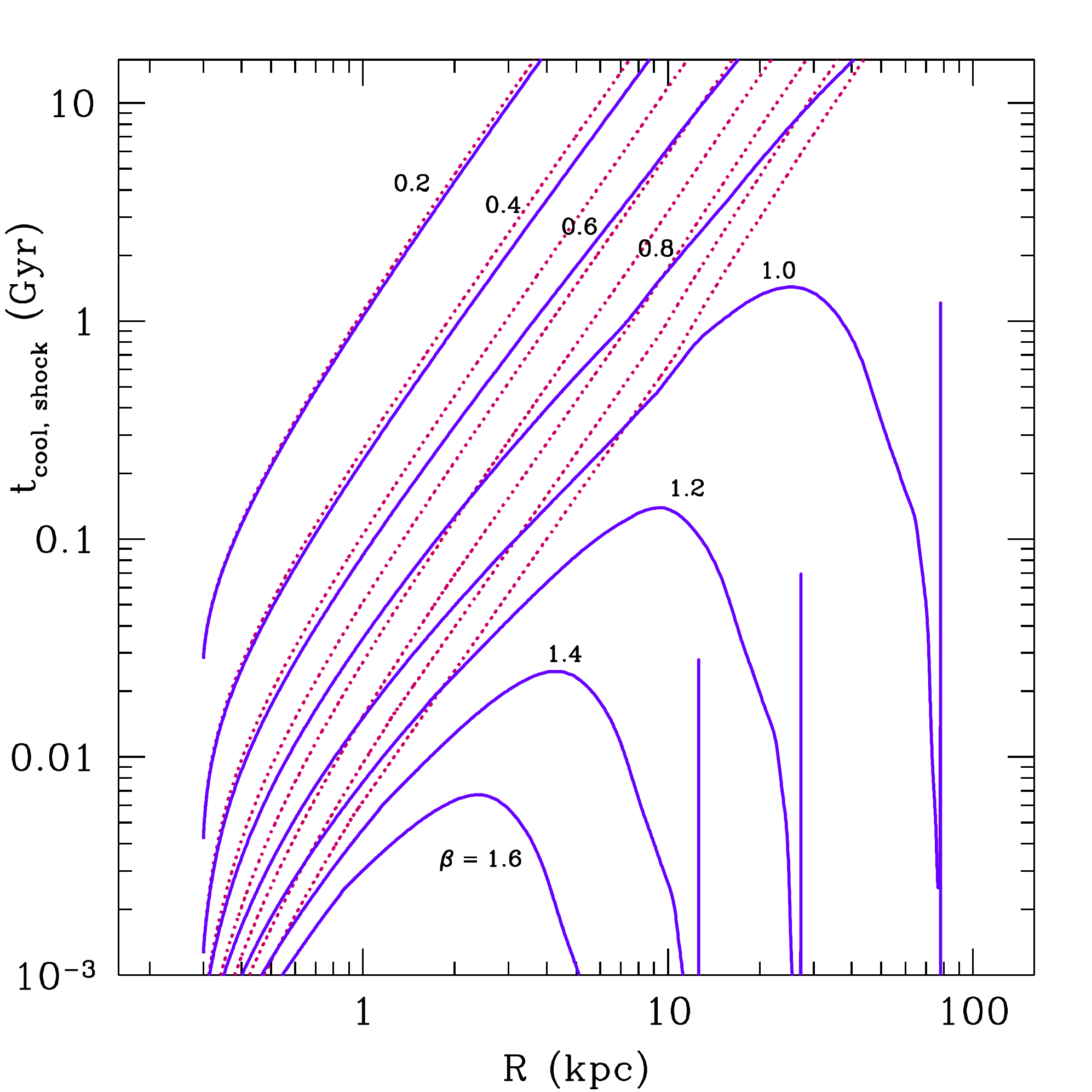}}
\caption{Cooling timescale for the wind material as a function of radius assuming that it goes through a strong shock, as might be appropriate for the reverse shock in a galactic wind-blown bubble propagating in the circumgalactic medium for the models presented in Figure \ref{figure:t}. The sharp decrease at the end of the high-$\beta$ profiles is from the shape of the profile just before the sonic point that develops in our solutions at large radii and reflects a breakdown of the time-steady assumption used throughout this work.}
\label{figure:ps}
\end{figure}

\subsection{The Fate of Cool Gas in Radiative Winds}
\label{section:halo}

Studies of quasar absorption lines in the outskirts of foreground galaxies imply that massive halos of cool gas are prevalent at both high ($z\sim1-2$) and low ($z\sim0.2$) redshift \citep{steidel_cgm,tumlinson,rudie,werk}. Modeling the absorbing medium as $T\simeq10^4$\,K gas in photoionization equilibrium with a background metagalactic radiation field implies a total mass in this cool component that is comparable to or larger than the stellar mass of the central galaxy (see \citealt{werk} for a detailed discussion). Covering fractions of low-ionization material are high out to $\sim100-200$\,kpc around both star-forming and passive galaxies, while higher ionization OVI absorption is found more frequently around star-forming galaxies \citep{tumlinson}.  The observed linewidths are hundreds (not thousands) of km/s. The densities inferred from photionization modeling, $n\sim10^{-3}$\,cm$^{-3}$, are far too low for the gas to be in pressure equilibrium with an ambient medium at the halo virial temperature, which challenges an explanation based on thermal instability within a hydrostatic hot halo (\citealt{mo_miralda,maller_bullock}; see Fig.~12 of \citealt{werk}).

We suggest that the observed absorption may originate from rapidly cooling outflows of the type presented in this paper. The calculations in Figure \ref{figure:t} show that mass-loaded flows driven from rapidly star-forming galaxies can radiatively cool on $\gtrsim{\rm few}-20$\,kpc scale, that these flows can reach large distances (even assuming an isothermal gravitational potential well out to $200$\,kpc), and that they slow down from their initial $500-1000$\,km/s velocities as a result of the extended gravitational potential. On these scales we expect a multi-phase medium as a result of the non-linear development of the thermal instability at $r_{\rm cool}$ (Section \ref{section:ti}) and the interaction with intervening gas (Section \ref{section:rebirth}). It is notable that for $\beta=\dot{M}_{\rm hot}/\dot{M}_\star\gtrsim1$ we expect more material to be ejected than to be formed into stars, and that it is in precisely this regime that cooling should be strong (eq.~\ref{betamin}). For the examples shown in Figure \ref{figure:t}, the column density of photoionized $\sim10^4$\,K gas is $\sim10^{17}-10^{18}$\,cm$^{-2}$ on $\sim100$\,kpc scales for the no-gravity (red dotted) calculations with $\beta\gtrsim1$. For metallicities of $\sim0.1-1Z_\odot$, corresponding metal-line column densities would be in the observable regime. The blue solid lines evolve to high column density as they ``stop" at the sonic point, but here our time-steady calculations break down and more detailed time-dependent calculations are necessary to explore the dynamics. 

The low velocities and relatively high columns on large scales appear consistent with the observations, but at the cost of assuming that all of the systems probed have recently had strong mass-loaded galactic winds. This is prima facie problematic for the passive galaxy halos in \cite{werk}. However, a basic piece of wind physics might help explain why even early-type galaxies have wind-mass-loaded halos: galactic outflows do not expand into vacuum, as assumed in the calculations of Figure \ref{figure:t}. In fact, we expect a newly born, hot, fast galactic wind to expand into a pressurized circumgalactic medium (CGM), to sweep up and shock this material, forming a wind-driven bubble with a characteristic reverse and forward shock configuration. A sketch is shown in Figure \ref{figure:sketch}. In the simplest case,  a contact discontinuity separates the shocked wind material from the swept up and shocked CGM. Simple arguments show that a wind-driven bubble would reach pressure equilibrium with a {\it constant pressure} CGM on $\sim40-80$\,kpc scales, assuming a CGM density of $10^{-3}$\,cm$^{-3}$ and a temperature equal to the virial temperature. In a hydrostatic gas halo with a falling CGM pressure profile, the bubble may well run away to large scales, depending on the CGM structure, the wind lifetime, and its energy and ram pressure. Importantly, wind-driven bubbles expand slowly compared to the free wind velocities that drive them. There is thus a chance that the remnant bubble formed by a cooling galactic wind would persist on multi-hundred kpc scales for Gyr timescales, thus allowing cool gas absorption around galaxies that are passive today.

Such an explanation for cool halos at low-$z$ in the sample of \cite{werk} should be fully developed. An immediate objection to such a model would be that as the cooled wind encounters the reverse shock in the wind-driven bubble it will be shocked to high temperature, potentially shutting off cooling, and thus no low-temperature photoionized gas would be observed outside of the cooling wind region. Although our calculations do not address the issue of a wind-driven bubble directly, we can make a preliminary estimate of relevance to the problem by asking what the temperature and density in the reverse shock region {\it would be} if the wind profiles from Figure \ref{figure:t} shocked at any given radius.  If we assume a strong shockwave at any position, we can then compute the post-shock cooling timescale. Figure \ref{figure:ps} shows the result. We take the post-shock density and temperature to be $n_{\rm RS}(r)=4n_{\rm wind}(r)$ and $T_{\rm RS}(r)=(3/16)v_{\rm wind}^2(r)m_p/k_B$ and we then calculate the cooling time 
\beq 
t_{\rm cool,\,shock}=k_B T_{\rm RS}/n_{\rm RS}\Lambda(T_{\rm RS}).
\label{tshock}
\eeq
 The models with and without gravity differ significantly. The high-$\beta$ models without gravity (red dotted) have cooling times of order Gyr only on scales below $\sim10$\,kpc. As the material goes through the reverse shock, it is heated to high temperature where the cooling time is long. The high post-shock temperature follows directly from the high wind velocities that are maintained even after cooling in the no-gravity models.  In addition, at large radius these models shock at very low wind density (see left middle panel Fig.~\ref{figure:t}), which further prevents cooling. 
 
In contrast, the models with gravity (blue solid) slow down significantly, increasing the density relative to the models with no gravity, and have lower post-shock temperatures because of the lower wind velocity. Both effects act to decrease the cooling time. We see that for $\beta > 0.8$, all of the models have post-shock cooling timescales less than the Hubble time out to scales of $\sim100$\,kpc. These conclusions would likely be strengthened by calculating the bubble evolution together with the wind evolution since the relative velocity of the reverse shock and the wind fluid will lead to lower post-shock temperatures and shorter cooling times.

Hot galactic winds with $\beta\gtrsim1$ can thus populate the halos of galaxies with cool gas in two ways. First, the freely expanding wind cools on small scales, undergoes thermal instability, and then evolves on $100$\,Myr timescales to very large scales, as in Figure \ref{figure:t}. Second, these cool decelerating winds will shock on the ambient CGM, driving a wind-driven bubble, and although the post-shock cooling times can be long, they are generically less than the Hubble time. We thus expect the post-shock region to cool radiatively. See Figure \ref{figure:sketch}.

These pictures for the origin of cool gas in the halos of galaxies do away with the notion of pressure equilibrium with a virialized hot halo. They circumvent the problem of maintaining a large mass of cool material well below the virial temperature in the CGM for many halo dynamical times.

\section{Conclusions}
\label{section:conclusions}

Following \cite{wang95a,wang95b} and a number of other works including \citealt{efstathiou2000,silich2003,silich2004,tenorio,tenorio05,tenorio07,wunsch},  \citealt{breitschwerdt94,breitschwerdt99}, we investigate how initially hot adiabatic galactic winds cool radiatively on large scales.  We present approximate scaling relations for the cooling radius, density, column density, emission measure, and radiative efficiency (Section \ref{section:analytic}). Each of these quantities depends strongly on the hot wind mass loading parameter $\beta = \dot{M}_{\rm hot}/\dot{M}_\star$. However, hot winds generically undergo radiative instability --- in the sense that $t_{\rm cool}<t_{\rm adv}$ --- for $\beta\gtrsim0.5$ (eqs.~\ref{betamin} and \ref{betacrit}) with interesting and testable dependencies on the thermalization efficiency $\alpha$, the scale of the star forming region $R$, and the star formation rate $\dot{M}_\star$. In particular, as $R/\dot{M}_\star$ decreases, the $\beta$ threshold for the cooling instability decreases.  Among other implications, the relatively small range of $\beta$ for strong radiative cooling implies a well defined maximum velocity given by equation (\ref{vmax}) and a minimum column density given by equation (\ref{ncoolmin}), with weak dependencies on the star formation rate and host galaxy radius. 

Using the time-steady wind equations for an arbitrary heating and cooling function, we present numerical solutions showing the temperature, density, velocity, and column density as a function of radius in Figure \ref{figure:t}, both with and without an extended gravitational potential, and assuming PIE with a metagalactic UV background. Models with $\beta\gtrsim1$ (eq.~\ref{betamin}) undergo strong cooling on small scales with the temperature dropping by two orders of magnitude or more over a small fractional radial scale. The qualitative behavior of the solutions is largely independent of the metallicity of the gas and the character of the UV background. 

The cooled gas forms an extended photoionized region that can reach large scales, depending on the gravitational potential. Even in the absence of a surrounding medium, the radiatively cooled wind decelerates significantly, and in our models spontaneously develops a sonic point, which signals the breakdown of our time-steady assumption. Equation (\ref{rsonic}) shows that the deceleration profile is exponentially sensitive to the parameters of the wind and the velocity dispersion of the host galaxy. For the same values of $\beta$ and the thermalization efficiency $\alpha$, this strong dependence on  $\sigma$ means that cooled winds will be completely unbound from dwarf galaxies, but strongly bound to massive cluster galaxies. This physics may connect directly to the efficiency with which baryons are retained in halos of a given velocity dispersion, and thus inform discussions of the $z=0$ stellar mass function and halo mass function.

\begin{figure}
\centerline{\includegraphics[width=8.5cm]{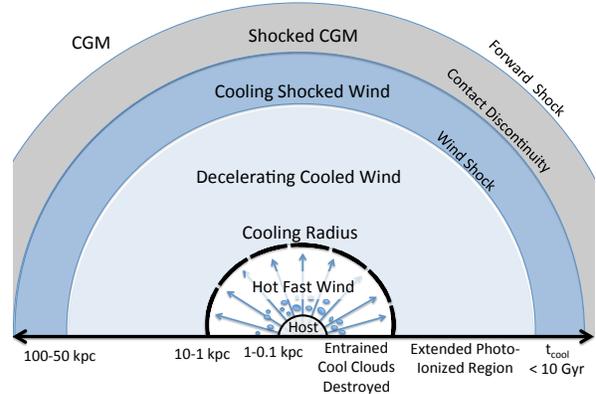}}
\caption{Sketch showing the basic evolutionary stages discussed and the three distinct cool gas components. The hot fast wind emerges from the host galaxy, accelerating cool clouds to small velocities before they are incorporated into the flow. The hot wind with high mass-loading factor cools radiatively at the cooling radius and undergoes thermal instability (Fig.~\ref{figure:t}). The cooled wind decelerates on large scales, forming an extended photoionized region. The wind then shocks on the inner edge of the wind-blown bubble driven into the surrounding CGM. The cooling timescale for the shocked wind can be significantly shorter than the Hubble time (Fig.~\ref{figure:ps}). The robust existence of multiple regions of cool gas in galactic winds and halos may explain its prevalence in observations of star-forming and passive galaxies. The sketch here resembles recent results from high-resolution galaxy formation studies (e.g., \citealt{onorbe}).}
\label{figure:sketch}
\end{figure}

The primary predictions of this model for cool gas are the scaling relations derived in Section \ref{section:winds}, combined with the basic picture that the hot gas cools via recombination, with a well-defined temperature progression and differential luminosity given by equation (\ref{lum}) and shown in Figure \ref{figure:dl}. This picture of strong radiative cooling may be a natural explanation for the high velocity line emission seen in local ULIRGs \citep{soto_fastlines}, as also emphasized by \cite{martin_lya}. Equations (\ref{sdscrit}) and (\ref{sds_R}) show that the critical star formation surface density required for the cooling radius to collapse to the scale of the star forming region is in the ULIRG range --- $\dot{\Sigma}_\star\simeq10^3$\,M$_\odot$ yr$^{-1}$ --- but that strong spatially extended radiative cooling is also expected from LBG-like galaxies with $\dot{\Sigma}_\star\gtrsim10$\,M$_\odot$ yr$^{-1}$. Strong radiative cooling may also provide an explanation for the exceptionally high velocity outflows probed by Mg II absorption in blue post-starburst galaxies \citep{tremonti,diamond,sell}. For example,  the system J0905$+$5759 reaches a velocity of $\simeq2500$\,km/s (Fig.~3 of \citealt{diamond}), close to $v_{\rm max}$ in equation (\ref{vmax}). Further tests of the model with ``down the barrel" absorption line studies should be undertaken. The application of the model to local starburst galaxies like NGC 253 and M82 is tentative and should be more fully explored with detailed models compared to high-resolution X-ray observations on small scales, as we highlight in Section \ref{section:coolwind}.

The rapid radiative cooling of nominally hot galactic winds may help explain observations at low and high redshift that show a significant amount of cool photoionized gas in the halos of galaxies (Section \ref{section:halo}). A particularly appealing aspect of this picture is the mass budget itself. The total mass in the cool gas halos is inferred to be comparable to or larger than the stellar mass of the galactic host \citep{werk}, which is in line with our prediction that $\beta=\dot{M}_{\rm hot}/\dot{M}_\star\gtrsim 0.5-1$ (eq.~\ref{betamin}) is required for strong radiative cooling.  

Importantly, there are two sources of cool gas in galaxy halos. The first is the cooled wind itself, the extended photoionized regions shown in Figure \ref{figure:t}. The second is in the reverse shock that must inevitably form as the wind sweeps up the surrounding circumgalactic medium. Figure \ref{figure:ps} shows that the post-shock cooling timescales can be less than the Hubble time on 100\,kpc scales if the wind decelerates. In this way the cool gas is reborn on larger scales and at later times as the bubble moves slowly outwards.

Combining this picture of the halo with our discussion of distributed mass-loading and ram pressure acceleration of cool clouds from Section \ref{section:rebirth}, we arrive at a story of two cool cloud transmigrations. Cool clouds within the host galaxy are first swept up by the ram pressure of the hot flow. These clouds are rapidly destroyed on small scales and at low velocities $\lesssim10^2$\,km/s by hydrodynamical instabilities, as is well-documented in both idealized high-resolution numerical simulations, and in parameterized models that take into account both the gravitational potential and the radial structure of the hot flow \citep{scannapieco,zhang15}. The hot flow, now with enhanced mass loading and density perturbations, cools radiatively on larger scales, forming an extended region of atomic and ionized gas moving at $\sim10^3$\,km/s (the first transmigration) and then decelerating in the host's extended gravitational potential. The high velocity cool wind then shocks on the surrounding circumgalactic medium, destroying the cool gas again, but because of the relatively short cooling times (Figure \ref{figure:ps}) the cool gas is again reborn on Gyr timescales (the second transmigration). In this way we generically expect three spatially-separated components of cool gas in rapidly star-forming galaxies, as illustrated in Figure \ref{figure:sketch}.

Many additional issues are left to be investigated. A pressing theoretical issue with direct observational implications is the character of the non-linear development of the thermal instability in a supersonic background. The medium is thermally unstable (Section \ref{section:ti}) and the initial density perturbations may grow by a factor of more than 100 at the cooling radius (eq.~\ref{nti}), but high resolution multi-dimensional simulations are needed in order to resolve the question of whether or not the entire medium cools monolithically or whether something akin to discrete cool clouds of a characteristic size precipitate out of the surrounding hot background. Our speculations on the fate of a cool wind-driven circumgalactic bubble also need to be confronted with both parameterized semi-analytic calculations and multi-dimensional numerical simulations with radiative cooling (e.g., \citealt{onorbe,sarkar}).

A number of microphysical issues also need to be investigated. We have shown that conduction (Section \ref{section:conduction}) can dominate the energy transport in low-$\beta$ outflows (eq.~\ref{betacond}), and our estimates suggest that the assumption of collisionality and thermal equilibrium for the ionized plasma break down. A consequence of this fact is that over essentially any range of $\beta$ the standard CC85 model is invalid. Comparing equation (\ref{betacond}) with equation (\ref{betamin}) we see that for $\beta\gtrsim0.5$ the breakdown of CC85 is a result of radiative cooling, while for $\beta\lesssim0.5$ the breakdown is a result of conduction.  Analytic and numerical investigations are needed to understand the temperature profile in the low-$\beta$ regime (eq.~\ref{parker}). These issues may well directly impact the interpretation of high-resolution X-ray observations of nearby starbursts like M82 \citep{strickland_m82}.

\section*{Acknowledgments}
TAT thanks the Kavli Institute for Theoretical Physics for support while preparing this work, and  Ondrej Pejcha, Crystal Martin, and Tim Heckman for discussions. TAT also thanks Smita Mathur and Laura Lopez for help in interpreting X-ray observations of local starburst galaxies.  We thank the Simons Foundation and the  organizers of the workshop {\it Galactic Winds: Beyond Phenomenology} (J.~Kollmeier and A.~Benson) where this work germinated.  We thank B.~Oppenheimer for providing the cooling/heating tables used in this work. We thank the anonymous referee for providing a timely, thorough, and helpful report. EQ was supported in part by NASA ATP grant 12-APT12-0183 and a Simons Investigator award from the Simons Foundation. TAT and DZ were supported in part by NASA Grant NNX10AD01G.  TAT was also supported by NSF Grant 1516967.


\appendix

\section{The Chevalier \& Clegg (1985) Solution}
\label{appendix:cc85}

The CC85 model neglects gravity and radiative cooling, and assumes uniform volumetric energy and mass injection rates inside the spherical starburst volume $r<R$ and adiabatic expansion for $r\ge R$. Assuming an ideal gas with adiabatic index $\gamma$, CC85 derived a self-similar solution for the Mach number ${\cal M}(r)$ of the form 
\beq
\left[\frac{\gamma-1+2/{\cal M}^2}{\gamma+1}\right]^{\frac{1+\gamma}{2(1+5\gamma)}}
\left[\frac{3\gamma+1/{\cal M}^2}{1+3\gamma}\right]^{-\frac{3\gamma+1}{5\gamma+1}}=\frac{r}{R}
\label{rltr}
\eeq
for $r<R$ and 
\beq
{\cal M}^{2/(\gamma-1)}\left[\frac{\gamma-1+2/{\cal M}^2}{\gamma+1}\right]^{\frac{1+\gamma}{2(\gamma-1)}}
=\left(\frac{r}{R}\right)^2
\label{bigr}
\eeq
for $r>R$.\footnote{See \cite{zhang14} for generalizations of equation (\ref{rltr}) to power-law energy- and mass-injection models for $r<R$.} Standard methods are used to solve these expressions for the Mach number at all radii. For $\gamma=5/3$, equation (\ref{bigr}) gives the useful expression
\beq
{\cal M}(r)^2\simeq16^{2/3}(r/R)^{4/3}\,\,\,\,\,({\cal M}^2\gg1).
\eeq
The overall normalization for the density, pressure, and temperature are given by the total mass injection rate $\dot{M}_{\rm hot}$ and the total energy injection rate $\dot{E}_{\rm hot}$ (eqs.~\ref{mdothot} and \ref{edothot}). The velocity of the flow is given by 
\beq
v^2 = \left[\frac{2{\cal M}^2}{{\cal M}^2+2/(\gamma-1)}\right]\frac{\dot{E}_{\rm hot}}{\dot{M}_{\rm hot}},
\eeq
the density follows from mass conservation ($\dot{M}_{\rm hot}=\Omega r^2\rho v$), and the gas temperature can be computed from the adiabatic sound speed via ${\cal M}$. For $r\gg R$, these are simply
\beq
\rho(r)\simeq\frac{1}{\Omega\sqrt{2}}\frac{\dot{M}_{\rm hot}^{3/2}}{\dot{E}_{\rm hot}^{1/2}}\frac{1}{r^2},
\eeq
\beq
v(r)^2\simeq v_\infty^2=\frac{2\dot{E}_{\rm hot}}{\dot{M}_{\rm hot}},
\eeq
and
\beq
T(r)\simeq \frac{2\dot{E}_{\rm hot}}{\dot{M}_{\rm hot}}\,\frac{m_p}{\gamma k_B\,{\cal M}^2}.
\eeq

\section{Derivation of the General Wind Equations}
\label{appendix:wind_equations}

The time-steady equations of hydrodynamics in spherical symmetry with optically-thin heating and cooling, and with only gravitational and pressure forces are
\beq
\dot{M}=4\pi r^2 \rho v={\rm constant},
\label{continuity}
\eeq
\beq
v\frac{dv}{dr}=-\frac{1}{\rho}\frac{dP}{dr}-\frac{GM(r)}{r^2},
\label{momentum}
\eeq
and
\beq
v\frac{d\epsilon}{dr}-\frac{vP}{\rho^2}\frac{d\rho}{dr}=\dot{q},
\label{energy}
\eeq
where $\dot{q}$ is the net heating/cooling per gram, $v$ is the velocity, $\rho$ is the mass density, $r$ is the radial coordinate, $M(r)$ is the enclosed mass, $\epsilon$ is the specific internal energy, and $P$ is the pressure. Writing equation (\ref{continuity}) in differential form as 
\beq
\frac{d\ln\rho}{dr}+\frac{d\ln v}{dr}+\frac{2}{r}=0
\eeq
and employing the thermodynamic identities 
\beq
dP=\left.\frac{\partial P}{\partial T}\right|_\rho\,\delta T+\left.\frac{\partial P}{\partial \rho}\right|_T\,\delta \rho
\eeq
and
\beq
d\epsilon=\left.\frac{\partial \epsilon}{\partial T}\right|_\rho\,\delta T+\left.\frac{\partial \epsilon}{\partial \rho}\right|_T\,\delta \rho
\eeq
to expand the radial pressure and specific energy derivatives, respectively, and then employing the identities
\beq
c_s^2=c_T^2+\frac{D^2}{C_V T}
\eeq
and 
\beq
\left.\frac{\partial\epsilon}{\partial \rho}\right|_T=-\frac{T}{\rho^2}\left.\frac{\partial P}{\partial T}\right|_\rho+\frac{P}{\rho^2},
\label{identity}
\eeq
where
\beq
c_s^2=\left.\frac{\partial P}{\partial \rho}\right|_s,\hspace*{.2cm}
c_T^2=\left.\frac{\partial P}{\partial \rho}\right|_T,
\eeq
\beq
C_V=\left.\frac{\partial \epsilon}{\partial T}\right|_\rho,\hspace*{.2cm}{\rm and}\,\,\,\,
D=\frac{T}{\rho}\left.\frac{\partial P}{\partial T}\right|_\rho,
\eeq
and substituting into equations (\ref{momentum}) and (\ref{energy}), we arrive at the three coupled ordinary differential equations given in equations (\ref{v})-(\ref{t}). Note that equation (\ref{identity}) is identically zero for an ideal gas, but the general expressions are recorded here for completeness. Also note that these expressions neglect changes in the chemical composition or chemical potential of the matter.


\begin{thebibliography}{}

\bibitem[Aguirre et al.(2001)]{aguirre01} Aguirre, A., Hernquist, 
L., Schaye, J., et al.\ 2001, \apj, 561, 521 

\bibitem[Balbus \& Soker(1989)]{bs89} Balbus, S.~A., \& Soker, N.\ 1989, \apj, 341, 611 

\bibitem[Barcos-Mu{\~n}oz et al.(2015)]{barcos_munoz} Barcos-Mu{\~n}oz, L., Leroy, A.~K., Evans, A.~S., et al.\ 2015, \apj, 799, 10 

\bibitem[Breitschwerdt \& Schmutzler(1994)]{breitschwerdt94} Breitschwerdt, D., \& Schmutzler, T.\ 1994, \nat, 371, 774 

\bibitem[Breitschwerdt \& Schmutzler(1999)]{breitschwerdt99} Breitschwerdt, D., \& Schmutzler, T.\ 1999, \aap, 347, 650 

\bibitem[Chevalier \& Clegg(1985)]{cc85}Chevalier, R. A., \& Clegg, A. W. 1985, Nature, 317, 44

\bibitem[Cicone et al.(2014)]{cicone} Cicone, C., Maiolino, R., Sturm, E., et al.\ 2014, \aap, 562, A21 

\bibitem[Cooper et al.(2009)]{cooper09} Cooper, J.~L., Bicknell, 
G.~V., Sutherland, R.~S., \& Bland-Hawthorn, J.\ 2009, \apj, 703, 330 

\bibitem[Condon et al.(1991)]{condon91} Condon, J.~J., Huang, 
Z.-P., Yin, Q.~F., \& Thuan, T.~X.\ 1991, \apj, 378, 65 

\bibitem[Davis et al.(2014)]{davis_kt}  Davis, S.~W., Jiang, 
Y.-F., Stone, J.~M., \& Murray, N.\ 2014, \apj, 796, 107 

\bibitem[Dekel \& Silk(1986)]{dekel}Dekel, A., \& Silk, J. 1986, ApJ, 303, 39

\bibitem[Diamond-Stanic et al.(2012)]{diamond} Diamond-Stanic, 
A.~M., Moustakas, J., Tremonti, C.~A., et al.\ 2012, \apjl, 755, L26 

\bibitem[Downes \& Solomon(1998)]{downes_solomon} Downes, D., \& Solomon, P.~M.\ 1998, \apj, 507, 615 

\bibitem[Draine(2011)]{draine} Draine, B.~T.\ 2011, Physics of 
the Interstellar and Intergalactic Medium by Bruce T.~Draine.~Princeton 
University Press, 2011.~ISBN: 978-0-691-12214-4,  

\bibitem[Efstathiou(2000)]{efstathiou2000} Efstathiou, G.\ 2000, \mnras, 317, 697 

\bibitem[Erb(2008)]{erb08} Erb, D.~K.\ 2008, \apj, 674, 151 

\bibitem[Fabian(1994)]{fabian94} Fabian, A.~C.\ 1994, \araa, 32, 277 

\bibitem[Finlator \& Dav{\'e}(2008)]{finlator_dave} Finlator, K., \& Dav{\'e}, R.\ 2008, \mnras, 385, 2181 

\bibitem[F{\"o}rster Schreiber et al.(2001)]{forster2001} 
F{\"o}rster Schreiber, N.~M., Genzel, R., Lutz, D., Kunze, D., 
\& Sternberg, A.\ 2001, \apj, 552, 544 

\bibitem[F{\"o}rster Schreiber et al.(2003)]{forster2003} 
F{\"o}rster Schreiber, N.~M., Genzel, R., Lutz, D., 
\& Sternberg, A.\ 2003, \apj, 599, 193 

\bibitem[Geach et al.(2014)]{geach} Geach, J.~E., Hickox, 
R.~C., Diamond-Stanic, A.~M., et al.\ 2014, \nat, 516, 68 

\bibitem[Genzel et al.(2011)]{genzel_clump} Genzel, R., Newman, S., 
Jones, T., et al.\ 2011, \apj, 733, 101 

\bibitem[Grimes et al.(2006)]{grimes_vv114} Grimes, J.~P., Heckman, 
T., Hoopes, C., et al.\ 2006, \apj, 648, 310 

\bibitem[Haardt \& Madau(2012)]{haardt_madau} Haardt, F., \& Madau, P.\ 2012, \apj, 746, 125 

\bibitem[Hanasz et al.(2013)]{hanasz} Hanasz, M., Lesch, H., 
Naab, T., et al.\ 2013, \apjl, 777, L38 

\bibitem[Heckman et al.(1990)]{ham} Heckman, T.~M., Armus, 
L., \& Miley, G.~K.\ 1990, \apjs, 74, 833 

\bibitem[Hoopes et al.(2003)]{hoopes_OVI_M82} Hoopes, C.~G., Heckman, 
T.~M., Strickland, D.~K., \& Howk, J.~C.\ 2003, \apjl, 596, L175 

\bibitem[Hopkins et al.(2011)]{hopkins11_feedback} Hopkins, P.~F., 
Quataert, E., \& Murray, N.\ 2011, \mnras, 417, 950 

\bibitem[Hopkins et al.(2012)]{hopkins12_ism} Hopkins, P.~F., 
Quataert, E., \& Murray, N.\ 2012, \mnras, 421, 3488 

\bibitem[Hopkins et al.(2012)]{hopkins12_wind} Hopkins, P.~F., 
Quataert, E., \& Murray, N.\ 2012, \mnras, 421, 3522 

\bibitem[Jubelgas et al.(2008)]{jubelgas} Jubelgas, M., Springel, V., En{\ss}lin, T., \& Pfrommer, C.\ 2008, \aap, 481, 33 

\bibitem[Kennicutt(1998)]{kennicutt98} Kennicutt, R.~C., Jr.\ 1998, 
\apj, 498, 541 

\bibitem[Klein et al.(1994)]{klein} Klein, R.~I., McKee, C.~F., \& Colella, P.\ 1994, \apj, 420, 213 

\bibitem[Krumholz \& Thompson(2013)]{krumholz_thompson13} Krumholz, M.~R., \& Thompson, T.~A.\ 2013, \mnras, 434, 2329 

\bibitem[Lamers \& Cassinelli(1999)]{lamers_cassinelli} Lamers, H.~J.~G.~L.~M., \& Cassinelli, J.~P.\ 1999, Introduction to Stellar Winds, by Henny J.~G.~L.~M.~Lamers and Joseph P.~Cassinelli, pp.~452.~ISBN 0521593980.~Cambridge, UK: Cambridge University Press, June 1999.,  

\bibitem[Mac Low \& McCray(1988)]{maclow} Mac Low, M.-M., \& McCray, R.\ 1988, \apj, 324, 776 

\bibitem[Maller \& Bullock(2004)]{maller_bullock} Maller, A.~H., \& Bullock, J.~S.\ 2004, \mnras, 355, 694 

\bibitem[Marcolini et al.(2005)]{marcolini} Marcolini, A., Strickland, D.~K., D'Ercole, A., Heckman, T.~M., \& Hoopes, C.~G.\ 2005, \mnras, 362, 626 

\bibitem[Martin(2005)]{martin_2005} Martin, C.~L.\ 2005, \apj, 621, 227 

\bibitem[Martin et al.(2015)]{martin_lya} Martin, C.~L., Dijkstra, M., Henry, A., et al.\ 2015, \apj, 803, 6 

\bibitem[McCarthy et al.(1987)]{mccarthy} McCarthy, P.~J., van 
Breugel, W., \& Heckman, T.\ 1987, \aj, 93, 264 

\bibitem[McCourt et al.(2015)]{mccourt} McCourt, M., O'Leary, 
R.~M., Madigan, A.-M., \& Quataert, E.\ 2015, \mnras, 449, 2 

\bibitem[Mineo et al.(2012)]{mineo} Mineo, S., Gilfanov, M., 
\& Sunyaev, R.\ 2012, \mnras, 426, 1870 

\bibitem[Mo \& Miralda-Escude(1996)]{mo_miralda} Mo, H.~J., \& Miralda-Escude, J.\ 1996, \apj, 469, 589 

\bibitem[Murray et al.(2005)]{mqt} Murray, N., Quataert, 
E., \& Thompson, T.~A.\ 2005, \apj, 618, 569 

\bibitem[Murray et al.(2007)]{murray07} Murray, N., Martin, 
C.~L., Quataert, E., \& Thompson, T.~A.\ 2007, \apj, 660, 211 

\bibitem[Murray et al.(2010)]{mqt10} Murray, N., Quataert, 
E., \& Thompson, T.~A.\ 2010, \apj, 709, 191 

\bibitem[Murray et al.(2011)]{mmt} Murray, N., M{\'e}nard, 
B., \& Thompson, T.~A.\ 2011, \apj, 735, 66 

\bibitem[O{\~n}orbe et al.(2015)]{onorbe} O{\~n}orbe, J., 
Boylan-Kolchin, M., Bullock, J.~S., et al.\ 2015, arXiv:1502.02036 

\bibitem[Oppenheimer \& Dav{\'e}(2006)]{oppenheimer_dave06} Oppenheimer, B.~D., \& Dav{\'e}, R.\ 2006, \mnras, 373, 1265 

\bibitem[Oppenheimer \& Dav{\'e}(2008)]{oppenheimer_dave08} Oppenheimer, B.~D., \& Dav{\'e}, R.\ 2008, \mnras, 387, 577 

\bibitem[Oppenheimer \& Schaye(2013)]{oppenheimer_schaye} Oppenheimer, B.~D., \& Schaye, J.\ 2013, \mnras, 434, 1043 

\bibitem[Palou{\v s} et al.(2014)]{palous} Palou{\v s}, J., W{\"u}nsch, R., \& Tenorio-Tagle, G.\ 2014, \apj, 792, 105 

\bibitem[Parker(1964)]{parker64} Parker, E.~N.\ 1964, \apj, 139, 93
 
\bibitem[Peeples \& Shankar(2011)]{peeples_shankar} Peeples, M.~S., \& Shankar, F.\ 2011, \mnras, 417, 2962 

\bibitem[Proga et al.(2014)]{proga_cloud} Proga, D., Jiang, Y.-F., Davis, S.~W., Stone, J.~M., \& Smith, D.\ 2014, \apj, 780, 51 

\bibitem[Rieke et al.(1980)]{rieke} Rieke, G.~H., Lebofsky, 
M.~J., Thompson, R.~I., Low, F.~J., \& Tokunaga, A.~T.\ 1980, \apj, 238, 24 

\bibitem[Rudie et al.(2012)]{rudie} Rudie, G.~C., Steidel, C.~C., Trainor, R.~F., et al.\ 2012, \apj, 750, 67 

\bibitem[Rupke et al.(2005a)]{rupke_2005a} Rupke, D.~S., Veilleux, 
S., \& Sanders, D.~B.\ 2005, \apjs, 160, 87 

\bibitem[Rupke et al.(2005b)]{rupke_2005b} Rupke, D.~S., Veilleux, 
S., \& Sanders, D.~B.\ 2005, \apjs, 160, 115 

\bibitem[Rupke \& Veilleux(2011)]{rupke_mrk231} Rupke, D.~S.~N., \& Veilleux, S.\ 2011, \apjl, 729, L27 

\bibitem[Sakamoto et al.(2009)]{sakamoto} Sakamoto, K., Aalto, S., Wilner, D.~J., et al.\ 2009, \apjl, 700, L104 

\bibitem[Sarkar et al.(2015)]{sarkar} Sarkar, K.~C., Nath, 
B.~B., Sharma, P., \& Shchekinov, Y.\ 2015, \mnras, 448, 328 

\bibitem[Scannapieco \& Br{\"u}ggen(2015)]{scannapieco} Scannapieco, E., \& Br{\"u}ggen, M.\ 2015, arXiv:1503.06800 

\bibitem[Schneider \& Robertson(2014)]{schneider} Schneider, E.~E., \& Robertson, B.~E.\ 2014, arXiv:1410.4194 

\bibitem[Schure et al.(2009)]{schure} Schure, K.~M., Kosenko, D., Kaastra, J.~S., Keppens, R., \& Vink, J.\ 2009, \aap, 508, 751 

\bibitem[Scoville et al.(2014)]{scoville_arp220} Scoville, N., Sheth, 
K., Walter, F., et al.\ 2014, arXiv:1412.5183 

\bibitem[Sell et al.(2014)]{sell} Sell, P.~H., Tremonti, 
C.~A., Hickox, R.~C., et al.\ 2014, \mnras, 441, 3417 

\bibitem[Sharma et al.(2012)]{sharma} Sharma, P., McCourt, M., Quataert, E., \& Parrish, I.~J.\ 2012, \mnras, 420, 3174 

\bibitem[Shopbell \& Bland-Hawthorn(1998)]{shopbell} Shopbell, P.~L., \& Bland-Hawthorn, J.\ 1998, \apj, 493, 129 

\bibitem[Silich et al.(2003)]{silich2003} Silich, S., Tenorio-Tagle, G., \& Mu{\~n}oz-Tu{\~n}{\'o}n, C.\ 2003, \apj, 590, 791 

\bibitem[Silich et al.(2004)]{silich2004} Silich, S., Tenorio-Tagle, G., \& Rodr{\'{\i}}guez-Gonz{\'a}lez, A.\ 2004, \apj, 610, 226 

\bibitem[Socrates et al.(2008)]{socrates_cr} Socrates, A., Davis, 
S.~W., \& Ramirez-Ruiz, E.\ 2008, \apj, 687, 202 

\bibitem[Soto \& Martin(2012)]{soto_martin_diagnostic} Soto, K.~T., \& Martin, C.~L.\ 2012, \apjs, 203, 3 

\bibitem[Soto et al.(2012)]{soto_fastlines} Soto, K.~T., Martin, 
C.~L., Prescott, M.~K.~M., \& Armus, L.\ 2012, \apj, 757, 86 

\bibitem[Steidel et al.(2010)]{steidel_cgm} Steidel, C.~C., Erb, D.~K., Shapley, A.~E., et al.\ 2010, \apj, 717, 289 

\bibitem[Strickland et al.(1997)]{strickland_m82} Strickland, D.~K., Ponman, T.~J., \& Stevens, I.~R.\ 1997, \aap, 320, 378 

\bibitem[Strickland et al.(2000)]{strickland_ngc253a} Strickland, D.~K., 
Heckman, T.~M., Weaver, K.~A., \& Dahlem, M.\ 2000, \aj, 120, 2965 

\bibitem[Strickland et al.(2002)]{strickland_ngc253b} Strickland, D.~K., 
Heckman, T.~M., Weaver, K.~A., Hoopes, C.~G., \& Dahlem, M.\ 2002, \apj, 568, 689 

\bibitem[Strickland et al.(2004a)]{strickland_xa} Strickland, D.~K., 
Heckman, T.~M., Colbert, E.~J.~M., Hoopes, C.~G., 
\& Weaver, K.~A.\ 2004, \apjs, 151, 193 

\bibitem[Strickland et al.(2004b)]{strickland_xb} Strickland, D.~K., 
Heckman, T.~M., Colbert, E.~J.~M., Hoopes, C.~G., 
\& Weaver, K.~A.\ 2004, \apj, 606, 829 

\bibitem[Strickland  \& Heckman(2007)]{strickland_iron} Strickland, D.~K., \& Heckman, T.~M.\ 2007, \apj, 658, 258 

\bibitem[Strickland \& Heckman(2009)]{strickland_heckman} Strickland, D.~K., \& Heckman, T.~M.\ 2009, \apj, 697, 2030 

\bibitem[Suchkov et al.(1994)]{suchkov94} Suchkov, A.~A., 
Balsara, D.~S., Heckman, T.~M., \& Leitherer, C.\ 1994, \apj, 430, 511 

\bibitem[Suchkov et al.(1996)]{suchkov96} Suchkov, A.~A., Berman, 
V.~G., Heckman, T.~M., \& Balsara, D.~S.\ 1996, \apj, 463, 528 

\bibitem[Tenorio-Tagle et al.(2003)]{tenorio} Tenorio-Tagle, 
G., Silich, S., \& Mu{\~n}oz-Tu{\~n}{\'o}n, C.\ 2003, \apj, 597, 279 

\bibitem[Tenorio-Tagle et al.(2005)]{tenorio05} Tenorio-Tagle, 
G., Silich, S., Rodr{\'{\i}}guez-Gonz{\'a}lez, A., 
\& Mu{\~n}oz-Tu{\~n}{\'o}n, C.\ 2005, \apj, 620, 217 

\bibitem[Tenorio-Tagle et al.(2007)]{tenorio07} Tenorio-Tagle, 
G., W{\"u}nsch, R., Silich, S., \& Palou{\v s}, J.\ 2007, \apj, 658, 1196 

\bibitem[Thompson et al.(2005)]{tqm} Thompson, T.~A., 
Quataert, E., \& Murray, N.\ 2005, \apj, 630, 167 

\bibitem[Thompson \& Krumholz(2014)]{thompson_krumholz} Thompson, T.~A., \& Krumholz, M.~R.\ 2014, arXiv:1411.1769 

\bibitem[Thompson et al.(2015)]{thompson_2015} Thompson, T.~A., 
Fabian, A.~C., Quataert, E., \& Murray, N.\ 2015, \mnras, 449, 147 

\bibitem[Tremonti et al.(2007)]{tremonti} Tremonti, C.~A., 
Moustakas, J., \& Diamond-Stanic, A.~M.\ 2007, \apjl, 663, L77 

\bibitem[Tumlinson et al.(2011)]{tumlinson} Tumlinson, J., Thom, C., Werk, J.~K., et al.\ 2011, Science, 334, 948 

\bibitem[Veilleux et al.(2005)]{veilleux_review} Veilleux, S., Cecil, G., \& Bland-Hawthorn, J.\ 2005, \araa, 43, 769 

\bibitem[Veilleux et al.(2014)]{veilleux_mrk231} Veilleux, S., Teng, 
S.~H., Rupke, D.~S.~N., Maiolino, R., \& Sturm, E.\ 2014, \apj, 790, 116 

\bibitem[Wang(1995a)]{wang95a} Wang, B.\ 1995a, \apj, 444, 590 

\bibitem[Wang(1995b)]{wang95b} Wang, B.\ 1995b, \apjl, 444, L17 

\bibitem[Wang et al.(2014)]{wang_6240} Wang, J., Nardini, E., 
Fabbiano, G., et al.\ 2014, \apj, 781, 55 

\bibitem[Werk et al.(2014)]{werk} Werk, J.~K., Prochaska, J.~X., Tumlinson, J., et al.\ 2014, \apj, 792, 8 

\bibitem[Wiersma et al.(2009)]{wiersma} Wiersma, R.~P.~C., Schaye, J., \& Smith, B.~D.\ 2009, \mnras, 393, 99 

\bibitem[W{\"u}nsch et al.(2011)]{wunsch} W{\"u}nsch, R., Silich, S., Palou{\v s}, J., Tenorio-Tagle, G., \& Mu{\~n}oz-Tu{\~n}{\'o}n, C.\ 2011, \apj, 740, 75 

\bibitem[Wuyts et al.(2011)]{wuyts11} Wuyts, S., F{\"o}rster Schreiber, N.~M., van der Wel, A., et al.\ 2011, \apj, 742, 96 

\bibitem[Zhang \& Thompson(2012)]{zhang_thompson} Zhang, D., \& Thompson, T.~A.\ 2012, \mnras, 424, 1170 

\bibitem[Zhang et al.(2014)]{zhang14} Zhang, D., Thompson, 
T.~A., Murray, N., \& Quataert, E.\ 2014, \apj, 784, 93 

\bibitem[Zhang et al.(2015)]{zhang15}  Zhang, D., Thompson, 
T.~A., Quataert, E., \& Murray, N.\ 2015, arXiv:1507.01951 

\end{thebibliography}
\end{document}